\documentclass[10pt,journal,compsoc]{IEEEtran}

\usepackage{graphicx}
\usepackage{subfigure}
\usepackage{multirow}
\usepackage{color}
\usepackage{booktabs}
\usepackage{amsmath}
\usepackage{amsthm}
\usepackage{amsfonts}
\usepackage{bbm}
\usepackage{tabularx}
\newcolumntype{L}[1]{>{\raggedright\arraybackslash}p{#1}}
\newcolumntype{C}[1]{>{\centering\arraybackslash}p{#1}}
\newcolumntype{R}[1]{>{\raggedleft\arraybackslash}p{#1}}

\ifCLASSOPTIONcompsoc
  \usepackage[nocompress]{cite}
\else
  \usepackage{cite}
\fi
\ifCLASSINFOpdf
\else
\fi
\begin{document}
%
\title{User-specific Adaptive Fine-tuning for Cross-domain Recommendations~\thanks{Lei Chen and Fajie Yuan contribute equally.}}

\author{Lei~Chen,
        Fajie~Yuan,
        Jiaxi~Yang,
        Xiangnan~He,
        Chengming~Li,
        and~Min~Yang
\IEEEcompsocitemizethanks{
\IEEEcompsocthanksitem Lei Chen, Chengming Li and Min Yang are with Shenzhen Institute of Advanced Technology, Chinese Academy of Sciences, Shenzhen 518055, P.R. China.\protect
E-mail: lei.chen@siat.ac.cn, cm.li@siat.ac.cn and min.yang@siat.ac.cn.
\IEEEcompsocthanksitem Fajie Yuan is with Westlake University, Hangzhou 310024, P.R. China. A part of this work was finished when Fajie was AI researcher at Tencent Kandian Group.
\protect
E-mail: yuanfajie@westlake.edu.cn or fajieyuan@tencent.com.
\IEEEcompsocthanksitem Jiaxi Yang is with Huazhong University of Science and Technology, Wuhan 430074, P.R. China. \protect
E-mail: yangjiaxi@hust.edu.cn.
\IEEEcompsocthanksitem Xiangnan He is with University of Science and Technology of China, Hefei 230026, P.R. China. \protect
E-mail: xiangnanhe@gmail.com.
}
}

%
%

\markboth{IEEE TRANSACTIONS ON KNOWLEDGE AND DATA ENGINEERING, MANUSCRIPT}
{Shell \MakeLowercase{\textit{et al.}}: Bare Demo of IEEEtran.cls for Computer Society Journals}
%



\IEEEtitleabstractindextext{%
\begin{abstract}
Making accurate recommendations for cold-start users has been a longstanding and critical challenge for recommender systems (RS).
Cross-domain recommendations (CDR) offer a solution to tackle such a cold-start problem when there is no sufficient data for the users who have rarely used the system. 
An effective approach in CDR is to leverage the knowledge (e.g., user representations) learned from a related but different domain and transfer it to the target domain. 
Fine-tuning works as an effective transfer learning technique for this objective, which adapts the parameters of a pre-trained model from the source domain to the target domain. 
However, current methods are mainly based on the global fine-tuning strategy: the decision of which layers of the pre-trained model to freeze or fine-tune is taken for all users in the target domain.
In this paper, we argue that users in RS are personalized and should have their own  fine-tuning policies for better preference transfer learning. As such, we propose a novel User-specific Adaptive Fine-tuning method (UAF), selecting which layers of the pre-trained network to fine-tune, on a per user basis. Specifically, we devise a policy network with three alternative strategies to automatically decide which layers to be fine-tuned and which layers to have their parameters frozen for each user. 
Extensive experiments show that the proposed UAF exhibits significantly better and more robust performance for user cold-start recommendation. 
\end{abstract}

}

\maketitle

\IEEEdisplaynontitleabstractindextext

\IEEEpeerreviewmaketitle

\IEEEraisesectionheading{\section{Introduction}\label{sec:introduction}}
\IEEEPARstart{T}{he} past decade has seen a remarkable progress in deep learning (DL) and their applications in recommender systems (RS). A variety of neural network models~\cite{hidasi2015session,yuan2019simple,he2017neural,kang2018self,DL4Match} with larger and deeper architectures are proposed to model user interaction behaviors from online systems.
Among them, sequential recommendation models, such as the GRU4Rec~\cite{hidasi2015session}, NextItNet~\cite{yuan2019simple} and  SASRec ~\cite{kang2018self} have become especially popular since they in general require neither much feature engineering nor explicit user embeddings when making recommendations.
Despite the success, deep neural network models tend to fail in practice when their training data (i.e., user interactions) are insufficient.
Such scenarios widely exist in practical RS, when a large number of new users enroll in but have fewer interactions. Fortunately, the interaction behaviors of cold-start users are likely to be accessible from other online systems.
For example, a user in Amazon who has few purchase records might have hundreds of clicking interactions in YouTube. Such observed interaction feedback by YouTube could be a
 clue to infer her preference and make recommendations in 
Amazon.\footnote{In this paper, we assume data of both domains is available and put aside privacy concerns.} To this end,
cross-domain recommendations (CDR) that transfer knowledge from a related source domain, have been proposed and become a popular way to tackle the recommendation problem of cold users.

Transfer learning (TL) based on pre-training and fine-tuning \cite{yosinski2014transferable,radford2018improving,devlin2018bert,guo2019adafilter} is widely employed for domain adaptation tasks. Its basic idea is to first pre-train a large neural network model with  plenty of source data and then fine-tune it in the target domain where there might be insufficient training data.  Fine-tuning has become a core technique for successful TL and been widely studied in computer vision~\cite{rebuffi2018efficient} and NLP~\cite{devlin2018bert}; however, thus far,  it attracted very little attention in the recommendation field. In this paper, we aim to investigate how to 
perform fine-tuning for cross-domain recommendations
and encourage practitioners to apply our proposed fine-tuning as a common technique to further improve the performance of their CDR models. 

\begin{figure*}[t]
    \centering
    \includegraphics[scale=0.65]{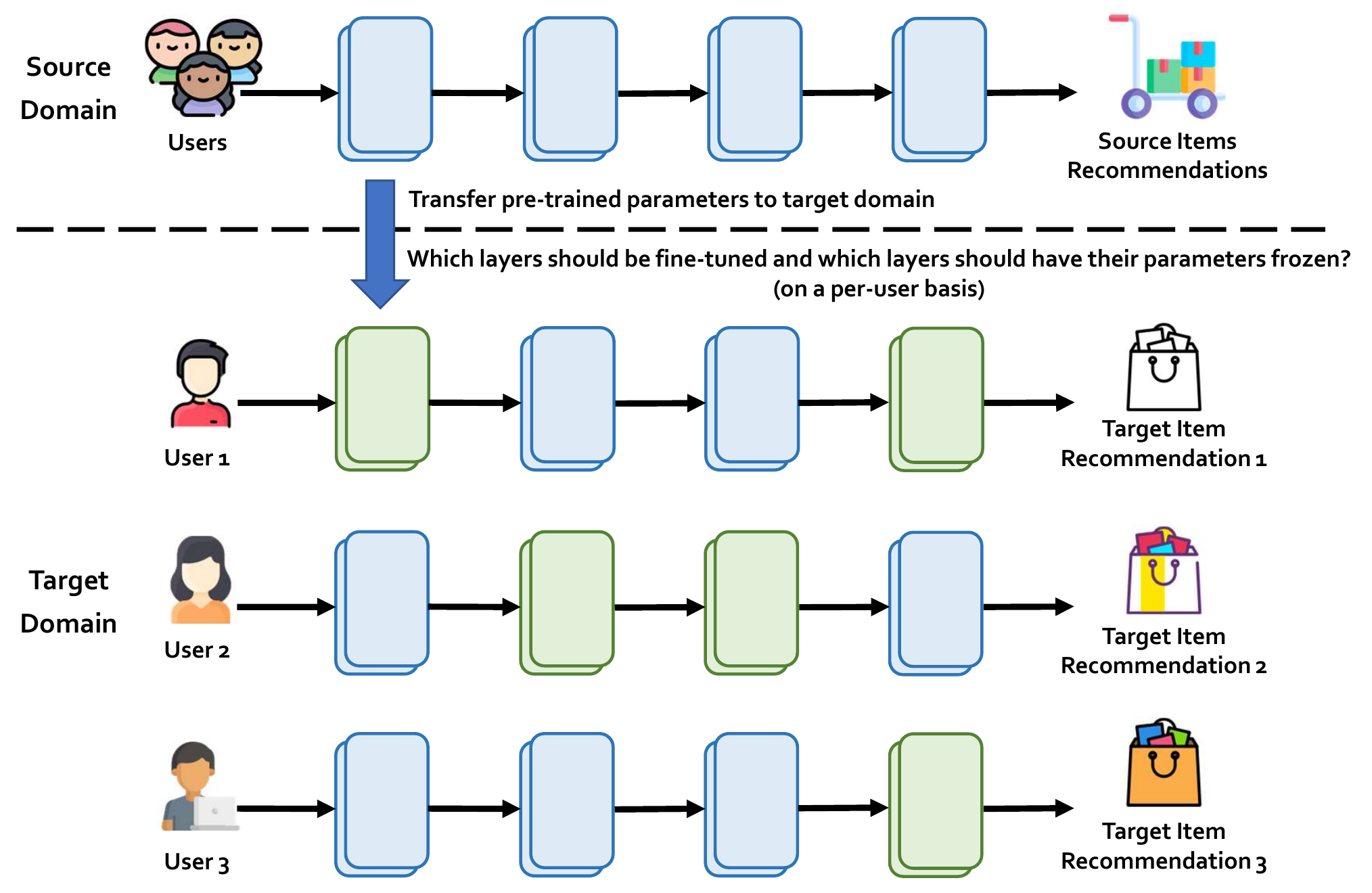}
    \caption{Fine-tuning layers on a per-user basis. The blue color represents the pre-trained layers, while the green color denotes the fine-tuned layers.}
    \label{fig:user}
\end{figure*}

There are several challenges when realizing the idea of fine-tuning of deep neural networks for CDR.  
(i) Great strides in accuracy have often been accompanied by increasing the depth and complexity of neural networks.  Fine-tuning such large models is prone to overfitting when the target domain has insufficient training data, and when all parameters are optimized.
(ii) Existing fine-tuning methods do not differentiate users --- i.e., treating the pre-trained parameters of each user equally in the target domain.  We argue that these methods restrict the power of the pre-trained  representations, because some users in the target domain may have high similarity with the source domain, s.t. routing these users through pre-trained parameters might be a better choice compared to updating them with limited target data. Ideally, for each individual user, we expect the  TL model is able to decide whether to use the pre-trained parameters or to fine-tune them, as illustrated in Figure~\ref{fig:user}.  For example,
user 1 requires the first and last layers to be fine-tuned, while the the two intermediate layers
are kept the same as their pre-trained model. On the other hand, user 3 requires more pre-trained parameters than user 1 and user 2 for the optimal recommendation accuracy in the target domain.

Motivated by the above-mentioned issues, in this paper we  propose UAF, a user-specific adaptive fine-tuning method which tailors fine-tuning for effective CDR. To begin with,
we first train an DL-based sequential recommendation model on a large source dataset and use it as our pre-trained model. This practice has been well verified by recent studies in ~\cite{yuan2020parameter,yuan2020one} (as well as in the NLP domains ~\cite{vaswani2017attention,radford2019language,brown2020language}) since sequential neural network can be trained in a self-supervised manner and thus could generate more universal user representations for effective transfer learning. 
In this paper, we choose the temporal CNN model NextItNet \cite{yuan2019simple} as the pre-trained backbone model given its efficient network structure~\cite{kalchbrenner2016neural,oord2016wavenet}, superior performance~\cite{wang2019towards,yuan2020future,yuan2019simple} and  widespread usage in modeling sequential recommendation data ~\cite{sun2020generic,yuan2020parameter,yuan2020future,yuan2020one,wang2020stackrec,chen2021user} in recent literature.
Then, we fine-tune the pre-trained model on the target dataset using UAF to address the cold user problem.
To be specific, for each user, we learn a policy network that makes binary decisions on whether 
to update or freeze the parameters of each layer in the pre-trained NextItNet.

The main contributions of this paper are as follows:
\begin{itemize}
 \item We propose a User-specific Adaptive Fine-tuning method for CDR on cold-start users. UAF allows each user in the target domain to have their own fine-tuning policy.
 To the best of our knowledge, UAF is the first `personalized' fine-tuning method designed for recommender systems.
 \item We propose a hard method, a soft method, and a RL method to derive the optimal fine-tuning policy without suffering from the non-differentiable problem of the discrete  policy decision functions. 
 \item Experimental results on four cross-domain datasets show that the proposed method achieves impressive results, especially when the training data in the target domain are insufficient. 
\end{itemize}

\section{Related Work}
We recapitulate important work for cross-domain recommendations. Since our pre-trained model is based on sequential recommendation model, we also briefly review some representative work based on deep learning (DL).
\subsection{Sequential Recommender Systems (SRS)}


In general,  DL-based  sequential recommendation (SR) models can be classified into three categories, namely RNN-based~\cite{hidasi2015session,tan2016improved},  CNN-based~\cite{tang2018personalized,yuan2019simple,yuan2020future} and self-attention~\cite{kang2018self,sun2019bert4rec} based methods.  
 Specifically, Hidasi \textit{et al.} \cite{hidasi2015session} proposed the first DL-based SR model GRU4Rec by adapting RNN  from language model in the NLP domain.
 Following this work, many extended RNN variants were proposed, which either optimized a new ranking loss~\cite{hidasi2018recurrent}, incorporated more context features~\cite{gabriel2019contextual}, or developed more advanced data augmentation~\cite{tan2016improved}. 
 While effective, these models rely heavily on the hidden states of the entire past, which cannot take full advantage of the parallel processing resources (e.g., GPU and TPU)~\cite{yuan2019simple} during training. Therefore, CNN and self-attention based models are proposed to mitigate such limitations, becoming more popular in recent literature~\cite{tang2018personalized,yuan2019simple,kang2018self,sun2019bert4rec}. 
 Among them, Tang \textit{et al.} \cite{tang2018personalized} proposed Caser, which embeds a sequence of user-item interactions into an ``image'' and learn sequential patterns as local features of the image by using wide convolutional filters. Subsequently,
\cite{yuan2019simple} proposed NextItNet, a  deep temporal CNN-based  recommendation model which particularly excels at modeling long-range item sequences. Later, GRec~\cite{yuan2020future} improved NextItNet by considering two directional context information. CpRec~\cite{sun2020generic} largely compressed NextItNet without hurting its recommendation accuracy.
StackRec~\cite{wang2019towards} and SkipRec~\cite{chen2021user} showed that NextItNet-style SR models could be stacked up to 100 layers for achieving its optimal accuracy, which distinguished from exiting work that usually 
applied less than 10 layers for evaluation. They also presented several methodologies to accelerate the training and inference processes of NextItNet. More recently, PeterRec~\cite{yuan2020parameter} and Conure~\cite{yuan2020one} demonstrated that the learned  user representations by NextItNet were generic and
could be transferred to solve various downstream recommendation problems.
Meanwhile, self-attention based  models, such as SASRec~\cite{kang2018self} and BERT4Rec~\cite{sun2019bert4rec}, also showed competitive performance for the SRS task. However, compared with NextItNet with linear complexity, these models often require quadratic time complexity to compute the self-attention matrix. Thereby, in this paper, we plan to use the well-known temporal CNN model (i.e., NextItNet) as the backbone network, given that the performance of which has been well evaluated by PeterRec~\cite{yuan2020parameter} and Conure~\cite{yuan2020one} for the CDR tasks~\cite{zeng2021knowledge}.

\subsection{Cross-domain Recommendations (CDR)}
While deep learning (DL) based models have shown impressive performance in providing personalized recommendations, they often suffer from the cold-start problem, when new users enroll in a system but have a few or no labeled training data~\cite{yuan2020parameter}. 
To deal with the cold-user problem, cross-domain recommendations (CDR) were proposed
and proved effective. 
CDR transfer user preference between domains based on similarity of users and items that occur in both domains~\cite{khan2017cross}.
CDR are particularly useful for recommendations of users who are cold in a target domain but have rich interactions from a source domain.
According to existing literature, CDR methods can be broadly divided into two types: joint learning based CDR, referred to as JLCDR, and two-stage (pre-training + fine-tuning) transfer learning based CDR, referred to as TLCDR.

In terms of JLCDR, ~\cite{kanagawa2019cross}  proposed a DL-based domain adaptation approach, domain separation network~\cite{bousmalis2016domain}, that solved the recommendation of cold-start users  by jointly learning a stacked denoising autoencoder. Similarly,  \cite{yuan2019darec} explored an autoencoder and adversarial training to extract abstract rating patterns that were shared for same users across domains. CoNet \cite{hu2018conet} is another representative cross-domain recommendation model using deep neural networks as the base model.  To enable effective knowledge transfer, CoNet introduced cross connections from the source network to the target and jointly trained objective functions of them.

Despite its effectiveness, JLCDR approaches are not efficient during training since the joint learning (e.g., multi-task learning) scheme is usually very costly.  In particular, to optimize the objective of the source domain is computationally very expensive as training data of it is often at a large-scale~\cite{devlin2018bert}. 
By contrast, TLCDR does not need to train the learner from scratch for target task.  Besides, compared with JLCDR, TLCDR can obtain better prediction accuracy~\cite{yuan2020one}. This is because the joint learning scheme of JLCDR
that trades off many objective functions
usually does not guarantee the optimal performance for the target  objective function~\cite{yuan2020parameter}. 

However,  existing literature of TLCDR  concentrated mainly on pre-training, i.e., developing  a more expressive or efficient base model~\cite{ni2018perceive,yang2019xlnet,devlin2018bert}.   
Closely related to this paper, a recent work called PeterRec~\cite{yuan2020parameter} firstly evidenced that the pre-training model based on self-supervised learning on user sequential behaviors largely improved the fine-tuning accuracy for the target task (i.e., CDR in our case).  Inspired by meta learning~\cite{bertinetto2016learning}, PeterRec injected a small-scale adapter network into the pre-trained model and optimized only the adapter during fine-tuning. The authors showed that PeterRec performed significantly better than fine-tuning the entire model when training data in the target domain is scarce.  In this paper, we argue that both PeterRec and the typical fine-tuning strategies achieve only sub-optimal accuracy since they perform fine-tuning globally for all users, ignoring the individual difference of users in  target tasks. 
To this end, we present an adaptive fine-tuning neural network architecture on a per-user basis. In other words, our proposed fine-tuning method is personalized --- each user is assigned to a different fine-tuning policy. To the best of our knowledge, UAF is the first attempt to the  CDR task using a personalized fine-tuning strategy.

This paper can be regarded as an extension of our previous conference paper published in AAAI 2021~\cite{chen2021user}. In \cite{chen2021user}, we propose a user-adaptive layer selection framework by learning to skip inactive hidden layers on a per-user basis to address the inference problem for traditional sequential recommender systems with super deep layers. In this paper, we target at cross-domain recommendations, which is a completely different research topic. We propose a user-specific adaptive fine-tuning method which tailors fine-tuning for effective cross-domain recommendations. In addition, we add a new UAF-Soft method by employing the weighted combination of freezed  pre-tained layers and  fine-tuned layers, which can preserve pre-trained information and receive valid fine-tuned information simultaneously.

\begin{figure*}[t]
    \centering
    \includegraphics[scale=1.0]{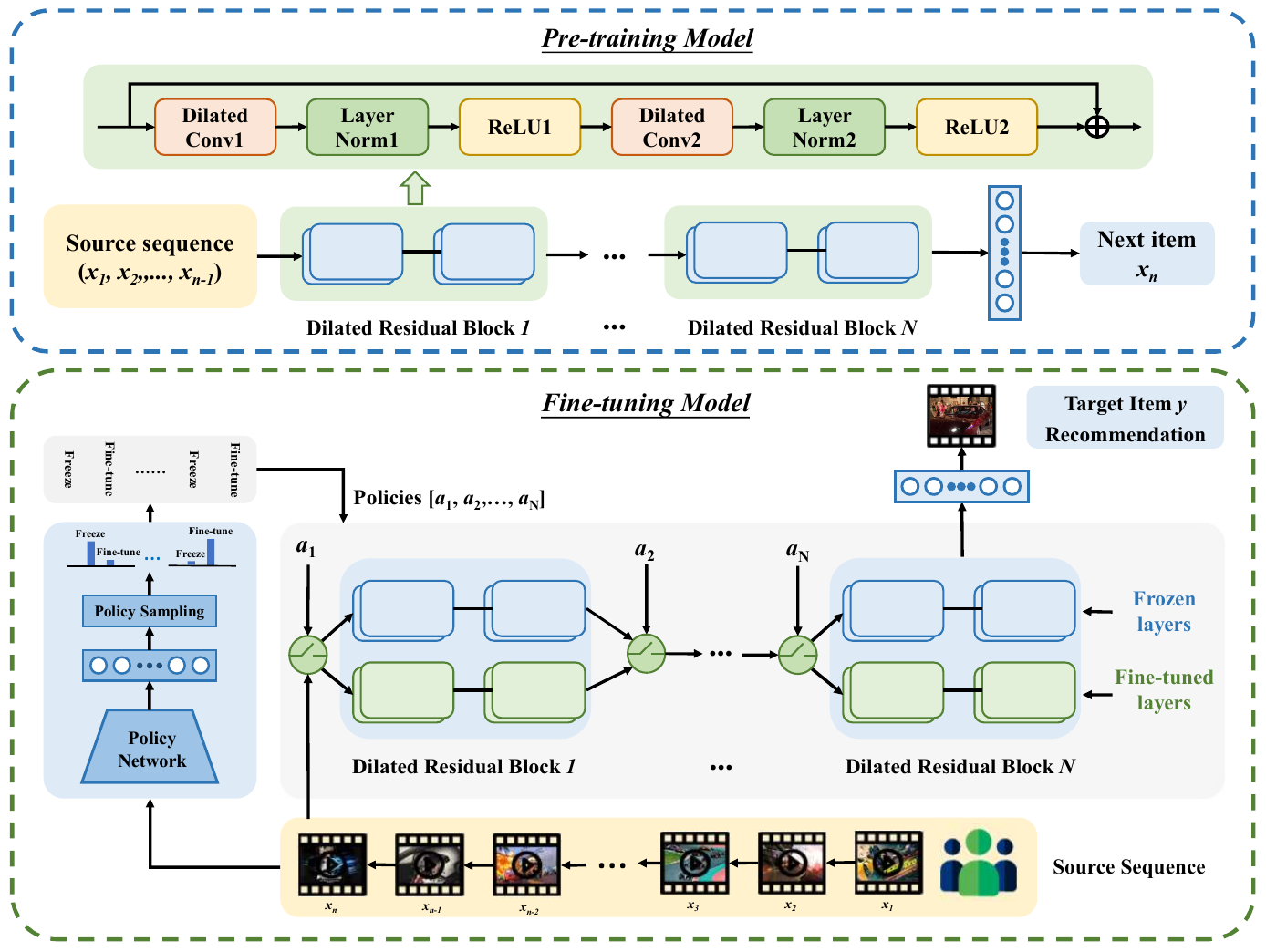}
    \caption{Model architecture. The proposed UAF consists of two modules: pre-training (above) and fine-tuning (below). In the pre-trained model, we learn the base network (NextItNet) on the source dataset to learn general user representations. In the fine-tuned model, we learn a user-specific adaptive fine-tuning strategy to automatically select reusable parameters of the pre-trained model on a per-user basis through a policy network.}
    \label{fig:model}
\end{figure*}

\section{Methodologies}
\subsection{Problem Definition}
Suppose that there are two domains: a source domain $\mathcal{S}$ with a large number of user interactions and a target domain $\mathcal{T}$ with limited user interactions (i.e.. the cold-user scenario). The input data takes the form of implicit feedback such as click logs and watching records.  Let $\mathcal{U}$ be the set of users shared in both domains. Each instance in $\mathcal{S}$ contains a userID $u \in \mathcal{U}$, and the user interaction sequence $x^{u}=\left\{x_{1}^{u}, \ldots, x_{n}^{u}\right\} \in \mathcal{X}$, where $x_{t}^{u}$ denotes the $t$-th interaction of $u$ and $\mathcal{X}$ (of size $\mathcal{|X|}$) is the set of items in source domain $\mathcal{S}$. Similarly, each instance in $\mathcal{T}$ consists of a userID $u$ and item (label) $y \in \mathcal{Y}$, i.e., $(u, y) \in \mathcal{T}$, where $\mathcal{Y}$ (of size $\mathcal{|Y|}$) is the set of items in target domain $\mathcal{T}$. 

Our objective is to recommend a set of items (a.k.a, the top-$N$ item recommendation) in the target domain to the user $u$ who has limited interactions but possesses sufficient sequential behaviors in the source domain. 

\subsection{The Overall Architecture}
Figure \ref{fig:model} illustrates the overall architecture of the proposed UAF method. The training procedure of UAF consists of two stages: pre-training and fine-tuning. In the first stage, we pre-train the base network (NextItNet) on the source dataset with  numerous user-item interactions. The pre-trained model is supposed to learn general user representations which are reusable for same users in the target dataset. Then, we propose a user-specific adaptive 
fine-tuning strategy to automatically select reusable layers from the pre-trained model on a per-user basis. 

\subsubsection{Pre-training Model}
As mentioned before, we choose NextItNet as the pre-trained model to learn the user representations given its superior performance in recent literature~\cite{wang2019towards,yuan2019simple}. 
NextItNet is composed of a stack of dilated convolutional layers, every two of which are wrapped by 
a residual block structure.
To be specific, each input item $x_i^u$ is converted into an embedding vector  $\mathbf{e}_i^u$, and the item sequence $x^u$ can be represented as an embedding matrix $\mathbf{E}^u= [\mathbf{e}_1^u \ldots \mathbf{e}_n^u]$. 
Then, we pass the item embeddings into a stack of dilated convolutional (DC) layers to learn the feature vector $\mathbf{E}^u_{l}$ which is expected to capture sequential dependencies.
Formally, the $l$-th residual block with the DC operation is formalized as:
\begin{equation}
\mathbf{E}^u_{l} =  \mathcal{F}_{l}(\mathbf{E}^u_{l-1}) + \mathbf{E}^u_{l-1}
\end{equation}
where $\mathbf{E}^u_{l-1}\in \mathbb{R}^{n\times k}$ and $\mathbf{E}^u_{l}\in \mathbb{R}^{n\times k}$ are the input and output of the $l$-th residual block considered, $k$ is the item embedding size. $\mathcal{F}_{l}(\mathbf{E}^u_{l-1})+ \mathbf{E}^u_{l-1}$ is a shortcut connection by element-wise addition. $\mathcal{F}_{l}(\mathbf{E}^u_{l-1})$ represents the residual mappings as:
\begin{equation}
\mathcal{F}_{l}(\mathbf{E}^u_{l-1}) =\sigma\left(\mathbf{L} \mathbf{N}_{2}\left(\psi_{2}\left(\sigma\left(\mathbf{L} \mathbf{N}_{1}\left(\psi_{1}(\mathbf{E}^u_{l-1})\right)\right)\right)\right)\right)
\end{equation}
where $\psi_1$ and $\psi_2$ represent the casual convolution operations. $\mathbf{L} \mathbf{N_1}$  and $\mathbf{L} \mathbf{N_2}$ represent layer normalization functions.  $\sigma$ is ReLU activation function. 
Finally, a softmax output layer is applied to predict the probability distribution for the next item $x^u_{n+1}$: 
\begin{equation}
p(x^u_{n+1}|x^u_{1:n}) = {\rm softmax}(\mathbf{W} \mathbf{E}^u_{l} + \mathbf{b})
\end{equation}
where $\mathbf{W}$ is the projection matrix, and $\mathbf{b}$ is the bias term. 

The joint probability $p\left(x^{u} ; \Omega\right)$ of each user sequence is computed by the product of the conditional distributions over the interacted items as follows:
\begin{equation}
p\left(x^{u} ; \Omega\right)=\prod_{i=1}^{n} p\left(x_{i}^{u} | x_{1:i-1}^{u}; \Omega\right)
\end{equation}
where $p\left(x_{i}^{u} | x_{1:i-1}^{u}; \Omega\right)$ is the predicted probability for the $i$-th item $x_{i}^{u}$ conditioned on all its previous interactions $\left\{x_{1}^{u}, \ldots, x_{i-1}^{u}\right\}$, and $\Omega$ is the set of parameters.

The objective function $\mathcal{G}(\mathcal{S} ; \Omega)$ of pre-trained NextItNet is to minimize the sum of negative log-likelihood of the joint probability, which is defined as:
\begin{equation}
\begin{aligned}
\mathcal{G}(\mathcal{S} ; \Omega) &=-\sum_{x^u \in \mathcal{S}} \log p\left(x^{u} ; \Omega\right) \\
&=-\sum_{x^u \in \mathcal{S}} \log \prod_{i=1}^{n} p\left(x_{i}^{u} | x_{1}^{u}, \ldots, x_{i-1}^{u} ; \Omega\right)
\end{aligned}
\end{equation}

\subsubsection{User-specific Adaptive Fine-tuning}
After the pre-training process, we employ UAF to identify which layers in the pre-trained model are required to be fine-tuned or freezed. 
For the $l$-th residual block in the pre-trained NextItNet model, in order to decide whether or not to fine-tune it during training, we first freeze the original block $\mathcal{F}_{l}$ and then create a new trainable block $\mathcal{\hat{F}}_{l}$, which is initialized with the parameters of $\mathcal{F}_{l}$. With the additional block $\mathcal{\hat{F}}_{l}$, the output of the $l$-th residual block in UAF is computed as:
\begin{equation}
\mathbf{E}^u_{l}=\mathbf{I}_{l}(\mathbf{E}^u) \mathcal{\hat{F}}_{l}\left(\mathbf{E}^u_{l-1}\right)+\left(1-\mathbf{I}_{l}(\mathbf{E}^u)\right) \mathcal{F}_{l}\left(\mathbf{E}^u_{l-1}\right)+\mathbf{E}^u_{l-1}
\label{eq_output}
\end{equation}
where $\mathbf{I}_{l}(\mathbf{E}^u)$ is a binary policy variable that indicates whether the residual block should be frozen or fine-tuned, conditioned on the input user sequence $\mathbf{E}^u$.
During training, given an input user sequence, the frozen block $\mathcal{F}_{l}$ trained on the source domain dataset is left unchanged, and the replicated block $\mathcal{\hat{F}}_{l}$ can be optimized towards the target domain dataset. 
Therefore, the input user sequence can either share the frozen block $\mathcal{F}_{l}$ or fine-tune the residual block $\mathcal{\hat{F}}_{l}$. $\mathbf{I}_{l}(\mathbf{E}^u)$ is sampled from a discrete distribution with two categories (freeze or fine-tune), which is parameterized by the output of a lightweight policy network. To be more specific, if $\mathbf{I}_{l}(\mathbf{E}^u)=0$, then the $l$-th frozen residual block is reused; otherwise, the $l$-th replicated residual block is fine-tuned.

By performing a series of convolution operations on frozen and fine-tuned layers with the input user sequence $\mathbf{E}^u$, we obtain the final hidden vector $\mathbf{h}_{n}$ of the last layer. Following~\cite{yuan2020parameter}, a fully-connected layer is applied to predict the score of item $y$ in the target domain by:
\begin{equation}
\mathbf{o}_y=\mathbf{W}_n \mathbf{h}_{n}+\mathbf{b}_n
\end{equation}
\noindent where $\mathbf{W}_n$ and $\mathbf{b}_n$ are parameters to be learned.

During the fine-tuninng stage, we adopt the popular pairwise ranking loss (BPR) \cite{rendle2009bpr} as the objective function $\mathcal{L}_{B P R}(\mathcal{T} ; \Theta)$ of UAF for top-$N$ recommendation: 
\begin{equation}
\mathcal{L}_{B P R}(\mathcal{T} ; \Theta)=-\sum_{(u, y) \in \mathcal{T}} \log (\delta \left(\mathbf{o}_{y^+}-\mathbf{o}_{y^-}\right))
\label{eq}
\end{equation}
\noindent where $\Theta$ is the parameters to be optimized, including 
pre-trained parameters (excluding the softmax matrix) and fine-tuned parameters (new softmax layer and newly added residual blocks),
$\delta$ is the logistic sigmoid function, $y^+$ is the positive label and $y^-$ is a negative label randomly sampled from $Y \backslash y^+$ following \cite{rendle2009bpr}, $\mathcal{T}$ contains all positive pairs $(u, y^+)$ and the selected negative pairs $(u, y^-)$ in the target dataset. 

\subsection{Policy Network}
To derive the optimal fine-tuning strategy given the input user sequence, we develop a policy network to output a binary policy vector, representing the actions to freeze or fine-tune each residual block in the pre-trained NextItNet. The policy network is implemented using a lightweight NextItNet model with dilations of \{1, 2, 4, 8\} (4 layers or 2 residual blocks). Without any restrictions, the policy network can also be implemented with other deep neural networks, e.g., a RNN model.
In this paper, we propose three alternative strategies (i.e., hard, soft and RL) to optimize the policy network. 

\begin{table*}
\centering
\caption{Statistics of the four experimental datasets. We reported the total number of instances and items in both source domain and target domain in each dataset. We regard the user sequence in source domain with each item in target domain of the same user as an instance of the target domain. Each user in the target dataset has several interactions, we form the training instances by using each interaction in the target dataset with the same user sequence in source domain. That is, if the target domain has 3 interactions, then we can generate 3 instances (maybe 2 for training and 1 for testing). We followed the same settings with previous work~\cite{yuan2020parameter}.}
{\renewcommand{\arraystretch}{1.2}
\begin{tabular}{C{1.8cm}|C{1.31cm}|C{1.31cm}|C{1.31cm}|C{1.31cm}|C{1.31cm}|C{1.31cm}|C{1.31cm}|C{1.31cm}}
\toprule
\multirow{2}{*}{\textbf{Domain}} & \multicolumn{2}{c|}{\textbf{ColdRec-1}}  & \multicolumn{2}{c}{\textbf{ColdRec-2}} & \multicolumn{2}{c|}{\textbf{ColdRec-3}}  & \multicolumn{2}{c}{\textbf{ColdRec-4}}  \\ \cline{2-9} 
& Instances & Items & Instances &  Items & Instances & Items & Instances &  Items \\ \midrule
$\mathcal{S}$              & 1,649,095        & 191,022      & 1,472,428        & 645,981   & 22,360        & 11,349      & 340,858        & 52,171     \\ \midrule
$\mathcal{T}$              & 3,798,114        & 20,343       & 2,947,688        & 17,880  & 24,643  & 162       & 1,267,117        & 6,288       \\ \bottomrule
\end{tabular}}
\label{tab:dataset}
\end{table*}

\subsubsection{UAF-Hard Method} 
Since the policy $\mathbf{I}_{l}(\mathbf{E}^u)$ is a discrete binary variable, it is intractable to optimize the policy network with backpropagation due to the non-differentiable problem. To resolve this issue, we propose using the Gumbeling Softmax sampling method~\cite{maddison2016concrete,jang2016categorical} to generate the actions (freeze or fine-tune) from a discrete distribution. We refer UAF with Gumbeling Softmax training as \textbf{UAF-Hard}. 


UAF-Hard draws samples $\mathbf{z}$ from a categorical distribution with class probabilities $\left\{\pi_{1}, \pi_{2}, \dots, \pi_{k}\right\}$. Here, we have $k=2$, indicating the freezing or fine-tuning actions. That is, for each residual block, $\pi_{1}$ and $\pi_{2}$ are scalars corresponding to the final output of the policy network.
\begin{equation}
\mathbf{z}=\text { one}_{-} \text {hot }\left(\arg \max _{i}\left[g_{i}+\log \pi_{i}\right]\right)
\label{eq1}
\end{equation}
\noindent where $\mathbf{z}$ is a one-hot vector and $\left\{g_{1}, g_{2}, \dots, g_{k}\right\}$ are i.i.d samples drawn from $Gumbel(0,1)$ distribution. The $Gumbel(0,1)$ distribution can be sampled using inverse transform sampling by drawing $u$ from a uniform distribution, i.e. $u \sim Uniform(0,1)$ and we can compute $g=-\log (-\log (u))$. 

The $\arg \max$ operation in Eq. (\ref{eq1}) is non-differentiable, but we can use the Gumbel Softmax distribution, which adopts softmax as a continuous relaxation to $\arg \max$ in order to allievate the non-differentiable problem. We relax the one-hot encoding of $\mathbf{z}$ to a real-valued vector $\boldsymbol{\alpha}$ using: 
\begin{equation}
\alpha_{i}=\frac{\exp \left(\left(\log \pi_{i}+g_{i}\right) / \tau\right)}{\sum_{j=1}^{k} \exp \left(\left(\log \pi_{j}+g_{j}\right) / \tau\right)} \quad \text { for } i=1, \dots, k
\label{eq2}
\end{equation}
\noindent where $\tau$ is a temperature parameter to control the discreteness of the output vector $\boldsymbol{\alpha}$. When $\tau$ becomes closer to 0, the samples from the Gumbel Softmax distribution become indistinguishable from the discrete distribution, i.e., almost the same as the one-hot vector~$\mathbf{z}$. Here we set $\tau$ to 10 by default.

Sampling the fine-tuned policy $\mathbf{I}_{l}(\mathbf{E}^u)$ from the Gumbel Softmax distribution allows us to backpropagate from the discrete binary decision samples to the policy network, as the Gumbel Softmax distribution is smooth for $\tau>0$ and has well-defined gradients with respect to the parameters $\pi_{i}$. Similar to \cite{wu2018blockdrop,guo2019spottune}, we generate all freezing/fine-tuning policies for all residual blocks at once for the trade-off of efficiency and accuracy. During the forward pass, we sample the fine-tuning policy $\mathbf{I}_{l}(\mathbf{E}^u)$  using Eq. (\ref{eq1}) for the $l$-th residual block. As for the backward pass, we approximate the gradients of the discrete samples by computing the gradients of the continuous softmax relaxation in Eq. (\ref{eq2}).

By using the above approach, we can easily achieve UAF-Hard method in a differentiable way and obtain the policy regarding which layers in pre-trained model should be fine-tuned. 
By using the objective function in Eq. (\ref{eq}) for the target domain, the policy network is jointly trained with the pre-trained NextItNet model in an end-to-end way.

\subsubsection{UAF-Soft Method} 
 In the UAF-Hard method, we learn a policy network to make binary decisions for fine-tuning or freezing the parameters of each layer in the pre-trained NextItNet model. That is, the decision $\mathbf{I}_{l}(\mathbf{E}^u)$ in every residual block is either 0 or 1. 
As an alternative to the UAF-Hard method, we propose a \textbf{UAF-Soft} method to learn the policy network by employing the weighted combination of freezed  pre-tained layers and  fine-tuned layers. In such a UAF-Soft method, the output of each residual block is no longer from  a certain layer, but from both pre-trained layers and fine-tuned layers. In this way,  we can preserve the useful pre-trained information and receive valid fine-tuned information simultaneously.


Specifically, we design a gate for each residual block to control the flow of parameters in the freezed pre-trained layers and fine-tuned layers in a soft way. The weight ratio in each gate is still obtained from the policy network using the lightweight NextItNet model with dilations of \{1, 2, 4, 8\}. We first put the interaction sequence from source domain into the policy network, and then pass it into the residual block to obtain the final hidden vector $\mathbf{h}_{p}$. Correspondingly, we will get   $\mathbf{v}$ for all residual blocks during the fine-tuning stage and adopt a logistic sigmoid function $\delta$ to achieve the weight distribution of the fine-tuning policy $\mathbf{I}_{l}(\mathbf{E}^u)$.

\begin{equation}
\mathbf{v}=\mathbf{W}_p \mathbf{h}_{p}+\mathbf{b}_p
\end{equation}

\begin{equation}
\mathbf{I}_{l}(\mathbf{E}^u) = \delta (\mathbf{v})
\end{equation}

where $\mathbf{W}_p$ is a projection matrix,  and $\mathbf{b}_p$ is a bias term.

Following UAF-Hard, we can obtain the output of the $l$-th residual block $\mathbf{E}^u_{l}$ at the fine-tuning stage using Eq. (\ref{eq_output}).
Since the sigmoid function is differentiable, we can directly train the policy network and the fine-tuning model in an end-to-end way.

\begin{table*}[t]
    \centering
    \caption{Overall results in terms of MRR@5 and HR@5 on all the four datasets. Note that the improvements of UAF over all baseline models on ColdRec-2/3/4 are statistically significant in terms of paired t-test with p-value $<$ 0.01.}
   {\renewcommand{\arraystretch}{1.2}
    \begin{tabular}{L{2.1cm}|C{1.31cm}|C{1.31cm}|C{1.31cm}|C{1.31cm}|C{1.31cm}|C{1.31cm}|C{1.31cm}|C{1.31cm}}
    \toprule
     \multirow{2}{*}{\textbf{Models}} & \multicolumn{2}{c|}{\textbf{ColdRec-1}} & \multicolumn{2}{c|}{\textbf{ColdRec-2}} & \multicolumn{2}{c|}{\textbf{ColdRec-3}} & \multicolumn{2}{c}{\textbf{ColdRec-4}} \\ \cline{2-9}
     & MRR@5 & HR@5 & MRR@5 & HR@5 & MRR@5 & HR@5 & MRR@5 & HR@5 \\
    \midrule
    Finetune-Zero  & 0.2332 & 0.4013 & 0.3632 & 0.5464 & 0.1174 & 0.2150  & 0.1929 & 0.3257 \\ 
    Finetune--CLS & 0.2419 & 0.4124 & 0.3712 & 0.5547 & 0.0909 & 0.1657 & 0.1935  & 0.3283 \\ 
    Finetune-Last1 & 0.2546 & 0.4307 & 0.3864 & 0.5724 & 0.1062 & 0.1918 & 0.2032 & 0.3433 \\ 
    Finetune-Last2 & 0.2575 & 0.4337 & 0.3885 & 0.5753 & 0.1176 & 0.2118 & 0.2073 & 0.3485 \\ 
    Finetune-All & 0.2565 & 0.4345 & 0.3880 & 0.5758 & 0.1185 & 0.2192 & 0.2077 & 0.3495 \\ 
    MTL & 0.2410 & 0.4114 & 0.3633 & 0.5468 & 0.1169 & 0.2181 & 0.2093 & 0.3527 \\ 
    PeterRec & 0.2578 & 0.4356 & 0.3886 & 0.5763 & 0.0993 & 0.1795 & 0.2117 & 0.3550 \\ 
    \midrule
    \textbf{UAF-Hard} & 0.2594 & 0.4378 & 0.3940 & 0.5824 & 0.1190 & \textbf{0.2212} & \textbf{0.2151} & \textbf{0.3640} \\  
    \textbf{UAF-Soft} & \textbf{0.2600} & \textbf{0.4397} & 0.3939 & \textbf{0.5841} & \textbf{0.1209} & 0.2189 & 0.2140 & 0.3613 \\ 
   \textbf{UAF-RL} & 0.2599 & 0.4386 & \textbf{0.3942} & 0.5837  & 0.1196 & 0.2197  & 0.2142 & 0.3619 \\
    \bottomrule
    \end{tabular}}
    \label{tab:result1}
\end{table*}

\subsubsection{UAF-RL Method} 
In fact, we can learn the policy network by employing the reinforcement learning algorithm (called \textbf{UAF-RL}) which is a popular technique for solving the non-differentiable problem.
The main idea is to learn the policy network that outputs the posterior probabilities of all the binary decisions for freezing or fine-tuning each block in the fine-tuning network. The policy network is trained using curriculum learning~\cite{DBLP:conf/slsp/Bengio13} to maximize a reward that incentivizes the use of as few blocks to fine-tune as possible while preserving the prediction accuracy. 
In this regard,  we can consider the potential trade-offs between computational cost and prediction accuracy. 

Let $\mathcal{A}  = \{0/1\}^N \in \mathbbm{R}^N$ denote the fine-tuning policies predicted by the policy network, where $N$ represents the number of residual blocks in the fine-tuning network. In particular, $\mathcal{A} \sim p(\pi_{i})$, where $\pi_i \in \{\pi_{1}, \pi_{2}\}$, indicating the freezing or fine-tuning actions. In order to evaluate the advantage of an action $\mathcal{A}_l$, we define the reward function as the following formula:
\begin{equation}
\mathbf{R}(\mathcal{A})=\left\{\begin{array}{ll}
1 - \left(\sum_l\mathbbm{1}(\mathcal{A}_l)/N\right)^{2} & \text { if correct } \\
-\gamma & \text { otherwise }
\end{array}\right.
\label{eq:reward}
\end{equation}
where $\mathbbm{1}(\cdot)$ is an indicator function, $\mathcal{A}_l$ represents the freezing or fine-tuning policy applied on the $l$-th block, $\gamma$ is a hyperparameter to penalize the wrong policy. The motivation behind the Eq. \ref{eq:reward} is two fold: (i) enabling the policy network to control the fine-tuning network so as to generate well-recommended items in target domain; (ii) achieving a significant computational cost reduction to meet the requirement of online service. Specifically, $\left(\sum_l\mathbbm{1}(\mathcal{A}_l)/N\right)^{2}$ measures the percentage of blocks that are fine-tuned. When a correct recommendation is produced, we incentivize block freezing by giving a larger reward to a policy that uses fewer blocks to fine-tune. In addition, we penalize incorrect recommendations with $\gamma$, which controls the trade-off between efficiency and effectiveness.  $\gamma$ is simply set to 1 in this paper.

To optimize the policy network, we adopt self-critical sequence training (SCST) \cite{DBLP:conf/cvpr/RennieMMRG17}, which is a form of the popular REINFORCE \cite{williams1992simple} algorithm, for model training. Rather than estimating a ``baseline" to normalize the rewards and reduce variance, SCST applies the output of its own test-time inference algorithm to normalize the rewards it experiences. In details, the exploration action $\mathcal{A}_l^s$ is obtained by sampling from the categorical distribution $p(\pi_{i})$ through modeling the source sequence, while the self-critical baseline is calculated by the greedy search, and the policy take action by $\hat{\mathcal{A}}_l = \mathop{\arg\max}\limits_{i} p(\pi_{i})$. To learn the optimal parameters of the policy network, we minimize the following SCST loss:
\begin{equation}
\mathcal{L}_{RL} = -\sum_{l=1}^{N} \log p(\mathcal{A}^s_l)\left(\mathbf{R}(\mathcal{A}^s_l) - \mathbf{R}(\hat{\mathcal{A}}_l)\right)
\end{equation}

\noindent where $p(\mathcal{A}^s_l)$ represents the probability to sample the exploration action $\mathcal{A}_l^s$. Since the self-critical baseline is based on the test-time estimate under the current model, SCST is forced to improve the performance of the model under the inference algorithm used at test time. At inference stage, we obtain the actual fine-tuning policy $\mathbf{I}_{l}(\mathbf{E}^u)$ by the greedy search $\mathop{\arg\max}\limits_{i} p(\pi_{i})$.

Finally, we jointly train the policy network and fine-tuning network in an end-to-end way and optimize the weighted-sum of the BPR loss and SCST loss in the UAF-RL method:
\begin{equation}
{\mathcal{L}}={\mathcal{L}}_{BPR}+\beta {\mathcal{L}}_{RL}
\end{equation}

\noindent where ${\mathcal{L}}_{BPR}$ and ${\mathcal{L}}_{RL}$ represent the standard BPR loss and the SCST loss, respectively, and $\beta$ is a hyperparameter to control the weight of the SCST loss, which is set to 1 in our experiments.

\section{Experimental Setup}
\subsection{Experimental Datasets}
To evaluate the effectiveness of UAF, we collect four transfer learning datasets from Tencent\footnote{https://www.tencent.com/en-us/}. Each of them is either from different recommender systems or
has different properties. Among them,  two datasets have  been made
 publicly available in \cite{yuan2020parameter}. The detailed properties of these datasets are described as follows.


\paragraph{ColdRec-1} 
ColdRec-1 includes both source and target domains. The source domain is the news recommendation dataset collected from QQ Browser\footnote{https://browser.qq.com}. Each instance is formed of a sequence of $n$ watching interactions of a user, where
$n$ is set to 50. Sequence length less than 50 will be padded with zero in the beginning, similar to ~\cite{yuan2020parameter}.
The target dataset is collected from Kandian\footnote{https://sdi.3g.qq.com/v/2019111020060911550}.  All users in Kandian are cold with at most 3 interactions (including clicks of news, videos or advertisements) and half of them have only one interaction. The source and target domains are connected by userID as all users
in the target dataset also have corresponding watching records in the source dataset.

\paragraph{ColdRec-2} It has similar characteristics with ColdRec-1 except that the source dataset contains 100 watching interactions for each user in the  sequence session.
In addition, the source dataset includes both news and video interactions. The target dataset has at most 5 interactions for each user. 

\paragraph{ColdRec-3} This is a private dataset collected by Tencent \textit{Oula} team. Similar to ColdRec-1, the source domain is the news recommendation dataset of QQ Browser, and the session length is set to 50. The target domain is an advertisement recommendation dataset, where 90\% users in target dataset have only one clicking action.

\paragraph{ColdRec-4} This is  also collected by Tencent \textit{Oula} team. The source dataset is collected from Tencent Video\footnote{https://v.qq.com/} including TV series, movies and short videos, and the session length is set to 50. The target dataset is collected from Kandian, and each user in target dataset has at most 5 clicks on news and videos.

The statistics are summarized in Table \ref{tab:dataset}.

\begin{table*}[t]
\centering
\caption{Results with a random policy (termed as UAF-Random) in terms of MRR@5 and HR@5 on all the four datasets.}
{\renewcommand{\arraystretch}{1.2}
\begin{tabular}{L{2.1cm}|C{1.31cm}|C{1.31cm}|C{1.31cm}|C{1.31cm}|C{1.31cm}|C{1.31cm}|C{1.31cm}|C{1.31cm}}
    \toprule
     \multirow{2}{*}{\textbf{Models}} & \multicolumn{2}{c|}{\textbf{ColdRec-1}} & \multicolumn{2}{c|}{\textbf{ColdRec-2}} & \multicolumn{2}{c|}{\textbf{ColdRec-3}} & \multicolumn{2}{c}{\textbf{ColdRec-4}} \\ \cline{2-9}
     & MRR@5 & HR@5 & MRR@5 & HR@5 & MRR@5 & HR@5 & MRR@5 & HR@5 \\
    \midrule
     UAF-Random & 0.2562 & 0.4346 & 0.3917  & 0.5796 & 0.1172 & 0.2158 & 0.2123 &  0.3587 \\ \midrule
     UAF-Hard & 0.2594 & 0.4378 & 0.3940 & 0.5824 & 0.1190 & \textbf{0.2212} & \textbf{0.2151} & \textbf{0.3640} \\  
     UAF-Soft & \textbf{0.2600} & \textbf{0.4397} & 0.3939 & \textbf{0.5841} & \textbf{0.1209} & 0.2189 & 0.2140 & 0.3613 \\ 
     UAF-RL & 0.2599 & 0.4386 & \textbf{0.3942} & 0.5837  & 0.1196 & 0.2197  & 0.2142 & 0.3619 \\ \bottomrule
\end{tabular}}
\label{tab:random_policy}
\end{table*}

\begin{table*}[t]
\centering
\caption{Results by replacing the policy networks (NextItNet vs. GRU) in terms of MRR@5 and HR@5 on all the four datasets.}
{\renewcommand{\arraystretch}{1.2}
\begin{tabular}{L{2.1cm}|C{1.31cm}|C{1.31cm}|C{1.31cm}|C{1.31cm}|C{1.31cm}|C{1.31cm}|C{1.31cm}|C{1.31cm}}
    \toprule
     \multirow{2}{*}{Models} & \multicolumn{2}{c|}{ColdRec-1} & \multicolumn{2}{c|}{ColdRec-2} & \multicolumn{2}{c|}{ColdRec-3} & \multicolumn{2}{c}{ColdRec-4} \\ \cline{2-9}
     & MRR@5 & HR@5 & MRR@5 & HR@5 & MRR@5 & HR@5 & MRR@5 & HR@5 \\
    \midrule
    UAF-Hard & 0.2594 & 0.4378 & 0.3940 & 0.5824 & 0.1190 & 0.2212 & 0.2151 & \textbf{0.3640} \\  
    UAF-Hard-GRU & 0.2591 & 0.4382 & \textbf{0.3943} & 0.5836 & 0.1186 & 0.2204 & \textbf{0.2159} & 0.3634 \\ \midrule 
    UAF-Soft & \textbf{0.2600} & \textbf{0.4397} & 0.3939 & 0.5841 & 0.1209 & 0.2189 & 0.2140 & 0.3613 \\ 
    UAF-Soft-GRU & 0.2598 & 0.4389 & 0.3941 & \textbf{0.5845} & \textbf{0.1217} & \textbf{0.2220} & 0.2151 & 0.3627 \\ \midrule  
    UAF-RL & 0.2599 & 0.4386 & 0.3942 & 0.5837  & 0.1196 & 0.2197  & 0.2142 & 0.3619 \\  
    UAF-RL-GRU & 0.2597 & 0.4381 & 0.3940 & 0.5833 & 0.1199 & 0.2201 & 0.2146 & 0.3625 \\ \bottomrule
\end{tabular}}
\label{tab:gru}
\end{table*}

\subsection{Baselines and Evaluation}
We compare our models (UAF-Hard, UAF-Soft and UAF-RL) with several powerful approaches. All baseline models share the same NextItNet architecture with ours. 

\begin{itemize}
\item \textbf{Finetune-Zero}: It is only trained on the target dataset without performing pre-training on the source dataset. That is, its parameters are randomly initialized. 

\item \textbf{Finetune-CLS}: It is pre-trained on the source dataset. We merely fine-tune the softmax layer of the model on the target dataset. 

\item \textbf{Finetune-Last-$N$} ($N=$1, 2): We pre-train the model on the source dataset and fine-tune the last $N$ residual blocks of the pre-trained network (with the softmax layer) on the target data following \cite{long2015learning}.

 
\item \textbf{Finetune-All}: We pre-train the model on the source dataset and fine-tune all parameters on the target dataset. Finetune-All is the most widely adopted fine-tuning approach, and usually performs better than Finetune-CLS and Finetune-Last-$N$ as the entire model is tuned~\cite{yuan2020parameter}.

\item \textbf{MTL}: We present a widely used multi-task learning (referred to as MTL) baseline by hard parameter sharing~\cite{caruana1997multitask}, similar to that used in \cite{yuan2020one}. MTL jointly learns the objective functions of both source and target domains rather than using the two-stage pre-training and fine-tuning framework. 

\item \textbf{PeterRec}: This is a recently proposed transfer learning framework which can be used to solve the cold-user problem~\cite{yuan2020parameter}. The basic idea is to fine-tune a small-sized adapter neural network inserted in the pre-trained network. We evaluate PeterRec by using the official code\footnote{https://github.com/fajieyuan/sigir2020\_peterrec} under our setting.
\end{itemize}


Our evaluation follows~\cite{yuan2020parameter} by adopting two popular top-$N$ metrics to measure the quality of recommended items, including MRR@$N$ (Mean Reciprocal Rank)~\cite{hidasi2015session} and HR@$N$ (Hit Ratio)~\cite{he2018adversarial}. Here $N$ is set to 5 for comparison. 

\begin{figure}[t]
    \centering
    \includegraphics[scale=0.6]{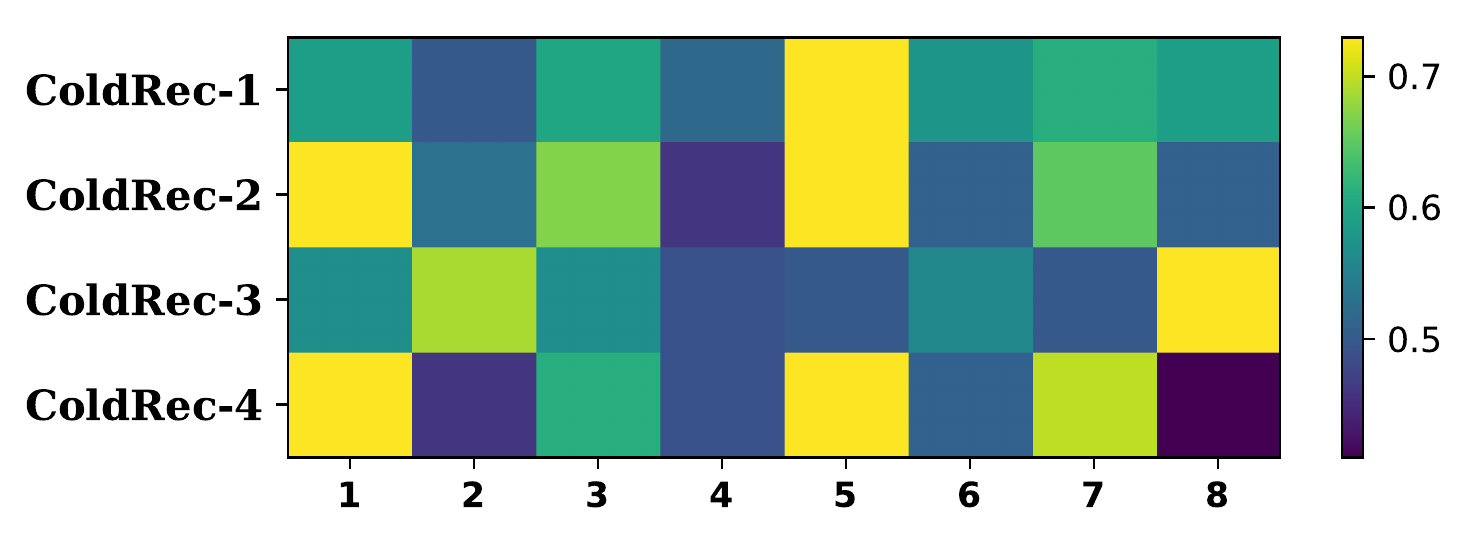}
    \caption{Visualization of the policy. x-axis denotes the residual blocks (i.e., every two CNN layers) from 1st to 8th.
    The color means the rate of utilization of the fine-tuned residual blocks for all users in average. i.e., yellow means a higher utilization rate, whereas blue denotes a lower rate.
    }
    \label{fig:heatmap}
\end{figure} 

\begin{table}[t]
    \centering
    \caption{Evaluation results with 10\% training data in terms of MRR@5 and HR@5 on ColdRec-1 and ColdRec-2.}
    {\renewcommand{\arraystretch}{1.2}
    \begin{tabular}{L{1.8cm}|C{1.2cm}|C{1.2cm}|C{1.2cm}|C{1.2cm}}
    \toprule
     \multirow{2}{*}{\textbf{Model}} & \multicolumn{2}{c|}{\textbf{ColdRec-1}} & \multicolumn{2}{c}{\textbf{ColdRec-2}} \\ \cline{2-5} 
     & MRR@5 & HR@5 & MRR@5 & HR@5 \\
    \midrule
    Finetune-Zero & 0.1624 & 0.2970 & 0.2766 & 0.4362  \\
    Finetune-All & 0.1857 & 0.3366 & 0.2915 & 0.4533 \\
    PeterRec & 0.1999 & 0.3562 & 0.2938 & 0.4557 \\ 
    \midrule
    \textbf{UAF-Hard} & 0.2120 & 0.3714 & \textbf{0.3025} & 0.4651 \\
    \textbf{UAF-Soft} & 0.2108 & 0.3704 & 0.3021 & 0.4649 \\
    \textbf{UAF-RL} & \textbf{0.2125} & \textbf{0.3721} & 0.3023 & \textbf{0.4655} \\
    \bottomrule
    \end{tabular}}
    \label{tab:result2}
\end{table}

\subsection{Implementation Details}
We randomly split each target dataset into training (80\%), validation (5\%) and test (15\%) sets. The grid search algorithm is applied on the validation sets to tune the hyperparameters. The size of each item embedding is set to be 128 for all the models.  The NextItNet architecture is implemented by using 16 dilation layers or 8 residual blocks (i.e., \{1, 2, 4, 8, 1, 2, 4, 8, 1, 2, 4, 8, 1, 2, 4, 8\}), which is used as the backbone of the pre-training and fine-tuning models. 
We employ Adam optimizer to train our models with learning rate $\eta$ of 0.0001. Batch size $b$ and kernel size are set to be 256 and 3, respectively. Regarding the policy network, we use dilations of \{1, 2, 4, 8\} (4 layers or 2 residual blocks), which is a lightweight neural network. Note that, as mentioned before, without any restrictions, the policy network can also be implemented with other deep neural networks, e.g., a RNN model. 
All the experiments are implemented in TensorFlow and trained on a single TITAN RTX GPU.
 
\begin{figure*}[t]
\begin{centering}
\includegraphics[width=0.9\columnwidth]{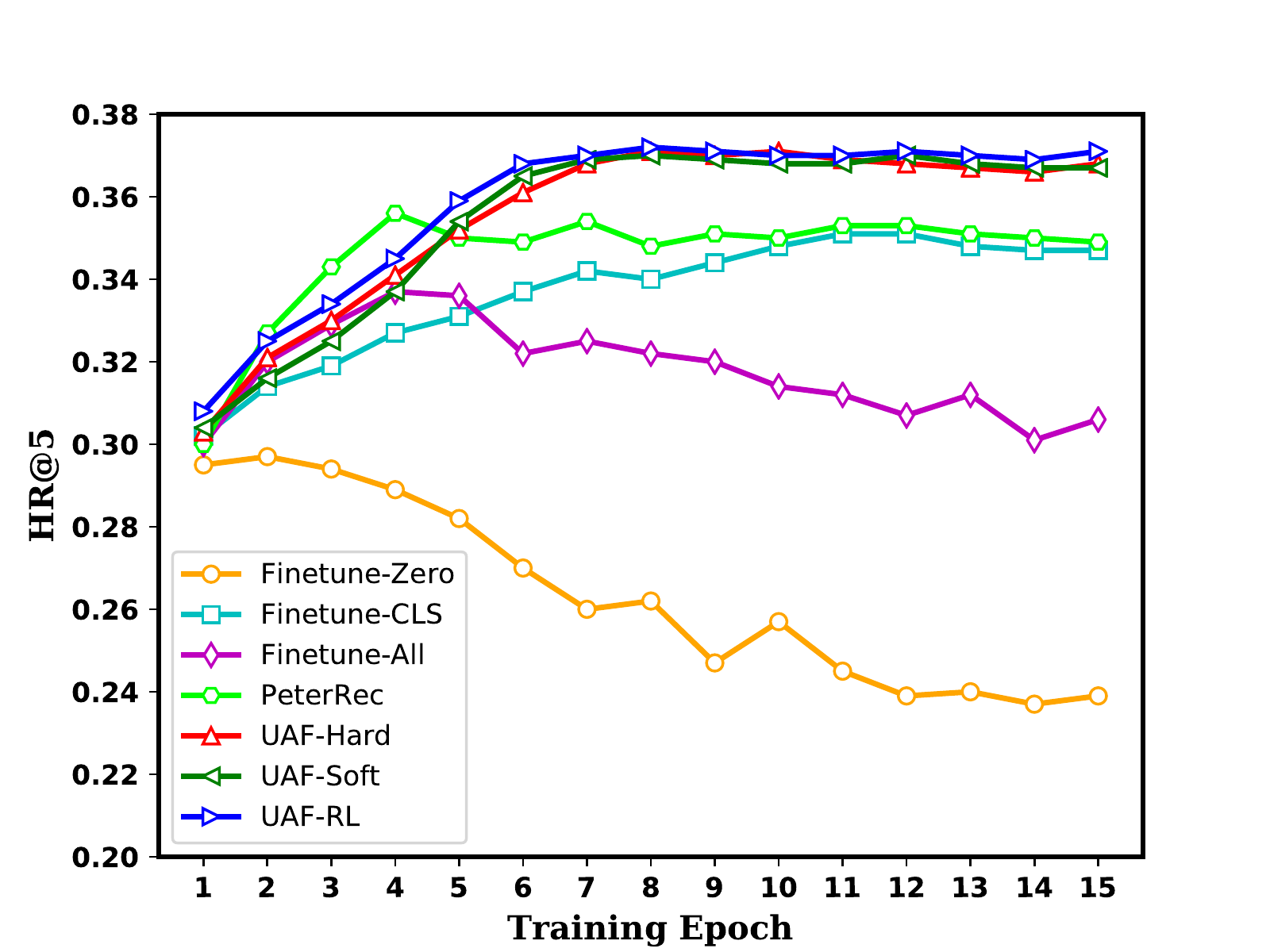}
\hspace{12pt}
\includegraphics[width=0.9\columnwidth]{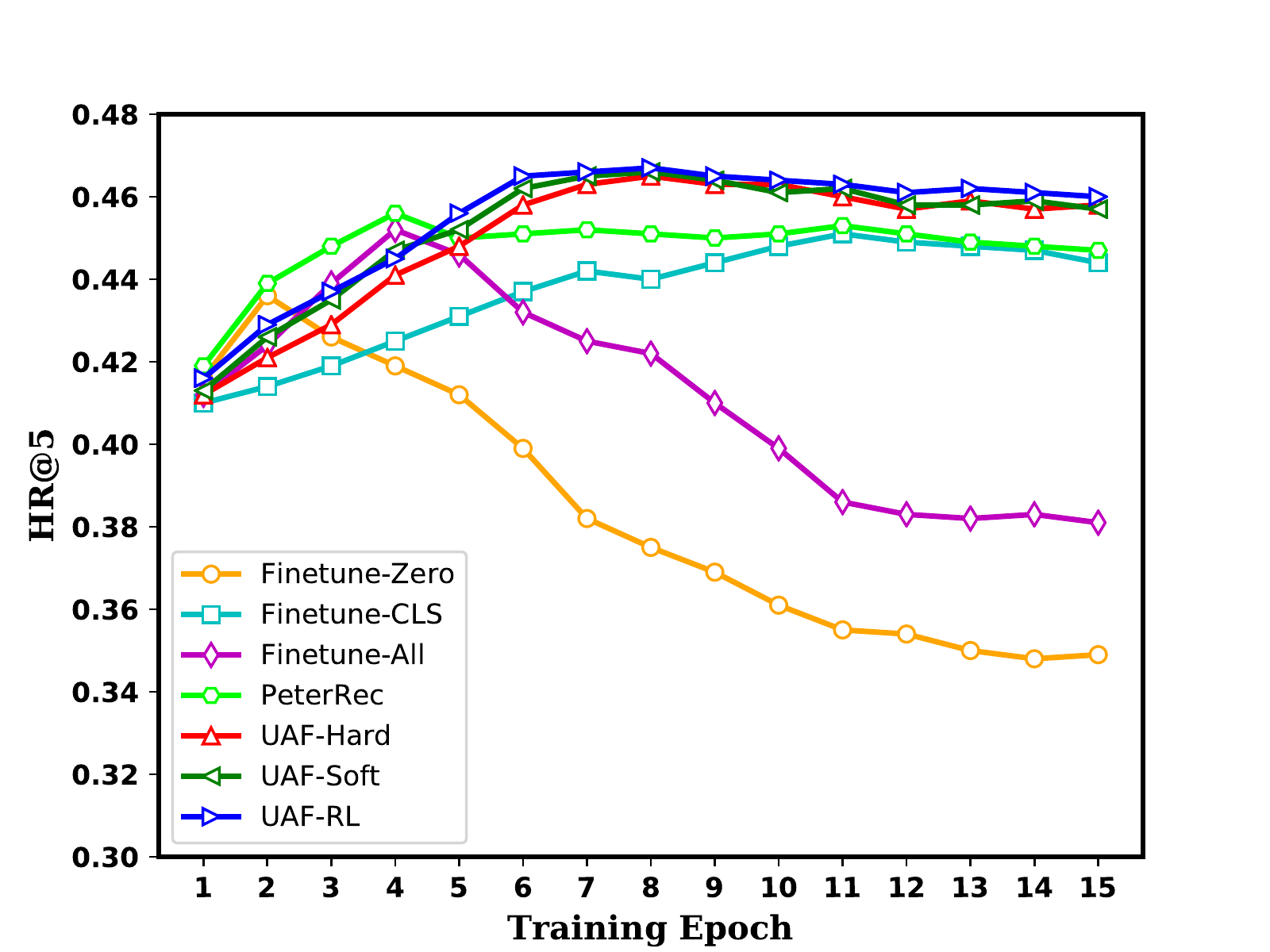}
\par\end{centering}
\caption{HR@5 curves for different models on ColdRec-1 and ColdRec-2.  
\label{fig:result}}
\end{figure*}

\section{Experimental Results}
As the key contribution of this work is to solve the CDR task with cold (or new) users through a user-specific adaptive fine-tuning method UAF, we evaluate UAF on four real-world datasets and conduct extensive experiments to answer the following research questions:

\begin{itemize}
    \item[(1)] \textbf{RQ1:} Whether the proposed UAF (i.e., UAF-Hard, UAF-Soft and UAF-RL) perform better than the existing fine-tuning approaches? 
    \item[(2)] \textbf{RQ2:} What are the effects of the policy network? Does it enable adaptive fine-tuning for each user?
    Can our policy network be replaced by other types of neural networks (e.g., a RNN model)?
    \item[(3)] \textbf{RQ3:} Can UAF stay robust when the training data in the target domain is insufficient?
    \item[(4)] \textbf{RQ4:} Is UAF a general method or does it work well in other tasks (e.g., user profile prediction)?
\end{itemize}



\subsection{Performance Comparison (RQ1)}
\label{rq2}
Table \ref{tab:result1} reports the overall results of UAF and baselines on the four datasets. We can make the following observations. First, Finetune-Zero performs worst among all baselines on ColdRec-1/2/4 in terms of both MRR@5 and HR@5. In particularly, by comparing  Finetune-Zero with Finetune-All on the three datasets, we can conclude that pre-training on the source domain is of great help to achieve better recommendation accuracy. On the other hand, Finetune-All obtains only 1\% improvement over Finetune-Zero on ColdRec-3. The result suggests that the correlation between the source and target domains on ColdRec-3  may be not as significant as ColdRec-1/2/4. The observation is reasonable since
advertisement recommendation may have some discrepancy with news recommendation in terms of user preference. Second, it is shown that Finetune-Last2 performs better than  Finetune-Last1, which further performs better than Finetune-CLS.
Our observations here are consistent with those in prior work. 
Third, MTL performs better than Finetune-Zero except on ColdRec-3, indicating the effectiveness of joint learning. Furthermore, Finetune-All outperforms MTL on ColdRec-1/2/3 but slightly underperforms it on ColdRec-4.  We argue that the optimal parameters learned for two or more objectives in MTL does not guarantee the optimal performance on the objective of the target domain.
By constrast, Finetune-All only cares about the objective on the target domain and thus performs generally better. Fourth, PeterRec achieves state-of-the-art results on ColdRec-1/2/4, but shows worse results on ColdRec-3. The result further indicates that the four datasets have different characteristics in terms of knowledge transfer: some datasets benefit a lot from pre-training, while others  may not. Again,
pre-training on ColdRec-3 is not as important as on other datasets, and as a result, PeterRec exhibits much worse performance because
all pre-trained parameters in PeterRec are not allowed to be modified.

Meanwhile, we observe that UAF, including UAF-Hard,  UAF-Soft and UAF-RL, performs obviously the best 
on ColdRec-2/3/4 and on par with PeterRec and FineAll on ColdRec-1. 
In fact, we further find that UAF performs significantly better than all other baselines even on ColdRec-1 with limited training data, as shown in Section~\ref{Robustness}.
We believe the
key advantage of UAF mainly comes from  its adaptive fine-tuning mechanism since these models are based on the same backbone network. 
By accomplishing the goal of per user per fine-tuning policy, UAF achieves more intelligent transfer learning across domains.



\subsection{Ablation Studies on the Policy Network (RQ2)}
To answer RQ2, we explore the following ablation studies to verify the effects of the policy network in UAF. Note that we may only report partial results for clarity if similar behaviors are observed.

\begin{table*}[t]
\centering
\caption{Statistics of the three user profile datasets. We reported the total number of instances and items in source domain, and instances and classes in target domain in each dataset. Note that $K=1000$ and $M=1000K$.}
{\renewcommand{\arraystretch}{1.2}
\begin{tabular}{C{2.0cm}|C{1.8cm}|C{1.8cm}|C{1.8cm}|C{1.8cm}|C{1.8cm}|C{1.8cm}}
\toprule
\multirow{2}{*}{\textbf{Domain}} & \multicolumn{2}{c|}{\textbf{GenEst}} & \multicolumn{2}{c|}{\textbf{AgeEst}} & \multicolumn{2}{c}{\textbf{LifeEst}} \\ \cline{2-7} 
& Instances &  Items$\&$Classes & Instances & Items$\&$Classes & Instances &  Items$\&$Classes \\ \midrule
$\mathcal{S}$ &   1.55M     & 646K & 1.55M        & 646K      & 1.08M        & 488K    \\ \midrule
$\mathcal{T}$ & 1.55M       & 2 & 1.55M   & 6  & 1.08M        & 8      \\ \bottomrule
\end{tabular}}
\label{tab:user_dataset}
\end{table*}

\begin{table*}[t]
\centering
\caption{Results in terms of Accuracy and F1 regarding user profile prediction.}
{\renewcommand{\arraystretch}{1.2}
\begin{tabular}{L{2.0cm}|C{1.8cm}|C{1.8cm}|C{1.8cm}|C{1.8cm}|C{1.8cm}|C{1.8cm}}
\toprule
\multirow{2}{*}{\textbf{Model}} & \multicolumn{2}{c|}{\textbf{GenEst}} & \multicolumn{2}{c|}{\textbf{AgeEst}} & \multicolumn{2}{c}{\textbf{LifeEst}} \\ 
\cline{2-7} & Accuracy & F1 & Accuracy & F1 & Accuracy & F1 \\
\midrule
 FinetuneCLS & 0.8882 & 0.7770 & 0.5717 & 0.2747 & 0.5165 & 0.2471 \\ 
 FinetuneAll & 0.9042 & 0.8205 & 0.6221  & 0.3457 & 0.5227 & 0.2670 \\ 
 PeterRec & 0.9040 & 0.8184 & 0.6163  & 0.3338 & 0.5342 & 0.2616 \\ \midrule
 \textbf{UAF-Hard} & \textbf{0.9051} & \textbf{0.8220} & 0.6243 & 0.3481 & \textbf{0.5491} & \textbf{0.2733}  \\
 \textbf{UAF-Soft} & 0.9048 & 0.8215  & \textbf{0.6259} & \textbf{0.3496} & 0.5484 & 0.2728  \\
 \textbf{UAF-RL} & 0.9049  & 0.8217 & 0.6250 & 0.3487 & 0.5478 & 0.2725   \\ \bottomrule
\end{tabular}}
\label{tab:user_results}
\end{table*}

\subsubsection{Random Policy}
In order to verify the effectiveness of the policy network, we compare it with a random policy network, termed as UAF-Random, and the results are shown in the Table~\ref{tab:random_policy}. The random policy network is implemented with a Bernoulli random distribution, which assigns 0.5 to the probability that the policy $\mathbf{I}_{l}(\mathbf{E}^u)$ equals 0 or 1. Note that our random seeds are kept fixed during training and inference.
Clearly, the results in  Table~\ref{tab:random_policy} show that a well-optimized policy outperforms a random policy. Interestingly, we find that UAF-Random performs comparably or slightly better than Finetune-All on the four datasets. The reason is because UAF-Random is able to combine parameters from both the pre-training and fine-tuning networks, albeit with a random way. However,
if we use evaluation seeds different from training, UAF-Random shows much worse results than Finetune-All.

\subsubsection{Visualization of Policies}
To better understand the fine-tuning policies of UAF, we visualize the utilization rates of the fine-tuned residual blocks in Figure \ref{fig:heatmap}. We only depict UAF-Hard since UAF-Soft and UAF-RL show similar behaviors. 
The illustration indicates that different datasets have completely different fine-tuning policies.
Clearly, the utilization rates of all fine-tuned residual blocks are always smaller than 1.
This indicates UAF does not only utilize fine-tuned blocks, instead, it also takes advantage
of the pre-trained blocks since the summation of  utilization rates of them equals to 1.
Hence, we conclude that UAF allows the fine-tuning model to automatically identify the right policy in determining which layers of the pre-trained model should be fine-tuned and which layers should have their parameters frozen on a per-user basis, which would be infeasible through a manual approach.

\subsubsection{Policy Network with GRU}
As mentioned before, the architecture of our policy network can be implemented with other deep neural networks without any restrictions. In order to verify the flexibility of it, we replace the original lightweight NextItNet with a Gated Recurrent Unit (GRU) for the policy network, termed as UAF-Hard-GRU, UAF-Soft-GRU and UAF-RL-GRU, respectively. We report the results in the Table \ref{tab:gru}. It clearly shows that designing the policy network by GRU yields comparable performance.

\subsection{Robustness Studies with Limited Training Data (RQ3)}
\label{Robustness}
Transfer learning is supposed to reduce the cost of labeling data. 
In this subsection, we investigate the effectiveness of UAF  with limited training samples. Specifically, we randomly choose 10\% of the  training examples in the target domain to fine-tune our models and the baselines. Table \ref{tab:result2} summarizes the results of all models on ColdRec-1 and ColdRec-2. As shown, our methods (including UAF-Hard, UAF-Soft and UAF-RL) achieve notably  better recommendation results  than the Finetune-All and PeterRec. 
For example, the UAF-Hard  method increases the MRR@5 metric by up  to  6.1\% on  ColdRec-1, compared with PeterRec. Therefore,
we conclude that the proposed UAF methods could be more powerful than these baselines when dealing with limited training data. 


It is well-known that deep models are prone to overfitting when training data is insufficient.
We here investigate the effectiveness of the proposed UAF methods against overfitting.  
We  depict the learning curves of different methods in Figure \ref{fig:result}. As shown,  we observe that UAF-Hard, UAF-Soft and UAF-RL prevent overfitting much better than the Finetune-All. For example, on ColdRec-1,  the HR@5 of the standard fine-tuning (Finetune-All) starts to decrease sharply after 5 epochs, whereas UAF-Hard, UAF-Soft and UAF-RL keep improving until 9 epochs, achieving  notably  higher HR@5. These results show that our methods can prevent overfitting better than many other models  since UAF utilizes both pre-trained and fine-tuned parameters, where pre-trained parameters will not suffer from the overfitting issue.

\subsection{Adaptability Experiments on User Profile Prediction Tasks (RQ4)}
User profiles are important features for generating accurate recommendations.
In~\cite{yuan2020parameter}, authors showed the transfer learning framework such as PeterRec and FineAll can also be used for predicting user profiles, which are of great help to lessen cold-start user problem in recommendation tasks. 
In this subsection, we investigate the versatility of UAF in the user profile prediction task.

Following \cite{yuan2020parameter}, we predict three types of user profile information (i.e., gender, age and life status). The statistics of the three user profile datasets are summarized in Table \ref{tab:user_dataset}. Specifically, each instance in GenEst is a user and his gender (male or female) label obtained by the registration information. Similar to GenEst, each instance in AgeEst is a user and his age bracket label - one class represents 10 years. And each instance in LifeEst is a user and his life status label, e.g., single, married, pregnancy or parenting. 

The results regarding the user profile prediction tasks are shown in Table \ref{tab:user_results}. We use classification accuracy (referred to as Accuracy) and F1 score (referred to as F1) as the evaluation metrics. Similar conclusions can be made as discussed in Subsection~\ref{rq2}. UAF with adaptive fine-tuning policies yields better results than baselines in all three datasets, which demonstrates the versatility of UAF while adapting it to different downstream tasks.

\section{Conclusion}
In this paper, we proposed a User-specific Adaptive Fine-tuning method (UAF) for the cross-domain recommendations. 
UAF introduces a policy network to automatically decide which layers of the pre-trained model should be fine-tuned and which layers should have their parameters frozen on a per-user basis. 
It combines the advantages of high-capacity pre-trained network and newly optimized fine-tuned network, leading to enhanced performance on the target task.
Extensive experiments on four real-world cross-domain recommendation datasets showed that UAF not only exhibited significantly better prediction accuracy, but also was more robust to overfitting, especially when the target domain has scarce training examples.

\section*{Acknowledgement}
Min Yang was partially supported by the National Natural Science Foundation of China (NSFC) (No. 61906185), Youth Innovation Promotion Association of CAS China (No. 2020357), Shenzhen Science and Technology Innovation Program (No. KQTD20190929172835662), 
Shenzhen Basic Research Foundation (No. JCYJ20200109113441941).

\ifCLASSOPTIONcaptionsoff
  \newpage
\fi


\bibliographystyle{IEEEtran}
\bibliography{ref.bib}

\begin{thebibliography}{10}
\providecommand{\url}[1]{#1}
\csname url@samestyle\endcsname
\providecommand{\newblock}{\relax}
\providecommand{\bibinfo}[2]{#2}
\providecommand{\BIBentrySTDinterwordspacing}{\spaceskip=0pt\relax}
\providecommand{\BIBentryALTinterwordstretchfactor}{4}
\providecommand{\BIBentryALTinterwordspacing}{\spaceskip=\fontdimen2\font plus
\BIBentryALTinterwordstretchfactor\fontdimen3\font minus
  \fontdimen4\font\relax}
\providecommand{\BIBforeignlanguage}[2]{{%
\expandafter\ifx\csname l@#1\endcsname\relax
\typeout{** WARNING: IEEEtran.bst: No hyphenation pattern has been}%
\typeout{** loaded for the language `#1'. Using the pattern for}%
\typeout{** the default language instead.}%
\else
\language=\csname l@#1\endcsname
\fi
#2}}
\providecommand{\BIBdecl}{\relax}
\BIBdecl

\bibitem{hidasi2015session}
B.~Hidasi, A.~Karatzoglou, L.~Baltrunas, and D.~Tikk, ``Session-based
  recommendations with recurrent neural networks,'' \emph{arXiv preprint
  arXiv:1511.06939}, 2015.

\bibitem{yuan2019simple}
F.~Yuan, A.~Karatzoglou, I.~Arapakis, J.~M. Jose, and X.~He, ``A simple
  convolutional generative network for next item recommendation,'' in
  \emph{WSDM}.\hskip 1em plus 0.5em minus 0.4em\relax ACM, 2019, pp. 582--590.

\bibitem{he2017neural}
X.~He, L.~Liao, H.~Zhang, L.~Nie, X.~Hu, and T.-S. Chua, ``Neural collaborative
  filtering,'' in \emph{WWW}, 2017, pp. 173--182.

\bibitem{kang2018self}
W.-C. Kang and J.~McAuley, ``Self-attentive sequential recommendation,'' in
  \emph{ICDM}.\hskip 1em plus 0.5em minus 0.4em\relax IEEE, 2018, pp. 197--206.

\bibitem{DL4Match}
J.~Xu, X.~He, and H.~Li, ``Deep learning for matching in search and
  recommendation,'' \emph{Foundations and Trends® in Information Retrieval},
  vol.~14, no. 2–3, pp. 102--288, 2020.

\bibitem{yosinski2014transferable}
J.~Yosinski, J.~Clune, Y.~Bengio, and H.~Lipson, ``How transferable are
  features in deep neural networks?'' in \emph{Advances in neural information
  processing systems}, 2014, pp. 3320--3328.

\bibitem{radford2018improving}
A.~Radford, K.~Narasimhan, T.~Salimans, and I.~Sutskever, ``Improving language
  understanding by generative pre-training,'' 2018.

\bibitem{devlin2018bert}
J.~Devlin, M.-W. Chang, K.~Lee, and K.~Toutanova, ``Bert: Pre-training of deep
  bidirectional transformers for language understanding,'' \emph{arXiv preprint
  arXiv:1810.04805}, 2018.

\bibitem{guo2019adafilter}
Y.~Guo, Y.~Li, L.~Wang, and T.~Rosing, ``Adafilter: Adaptive filter fine-tuning
  for deep transfer learning,'' \emph{arXiv preprint arXiv:1911.09659}, 2019.

\bibitem{rebuffi2018efficient}
S.-A. Rebuffi, H.~Bilen, and A.~Vedaldi, ``Efficient parametrization of
  multi-domain deep neural networks,'' in \emph{CVPR}, 2018, pp. 8119--8127.

\bibitem{yuan2020parameter}
F.~Yuan, X.~He, A.~Karatzoglou, and L.~Zhang, ``Parameter-efficient transfer
  from sequential behaviors for user modeling and recommendation,'' in
  \emph{Proceedings of the 43rd International ACM SIGIR Conference on Research
  and Development in Information Retrieval}, 2020, pp. 1469--1478.

\bibitem{yuan2020one}
F.~Yuan, G.~Zhang, A.~Karatzoglou, X.~He, J.~Jose, B.~Kong, and Y.~Li, ``One
  person, one model, one world: Learning continual user representation without
  forgetting,'' \emph{arXiv preprint arXiv:2009.13724}, 2020.

\bibitem{vaswani2017attention}
A.~Vaswani, N.~Shazeer, N.~Parmar, J.~Uszkoreit, L.~Jones, A.~N. Gomez,
  L.~Kaiser, and I.~Polosukhin, ``Attention is all you need,'' \emph{ICLR},
  2017.

\bibitem{radford2019language}
A.~Radford, J.~Wu, R.~Child, D.~Luan, D.~Amodei, and I.~Sutskever, ``Language
  models are unsupervised multitask learners,'' \emph{OpenAI blog}, vol.~1,
  no.~8, p.~9, 2019.

\bibitem{brown2020language}
T.~B. Brown, B.~Mann, N.~Ryder, M.~Subbiah, J.~Kaplan, P.~Dhariwal,
  A.~Neelakantan, P.~Shyam, G.~Sastry, A.~Askell \emph{et~al.}, ``Language
  models are few-shot learners,'' \emph{arXiv preprint arXiv:2005.14165}, 2020.

\bibitem{kalchbrenner2016neural}
N.~Kalchbrenner, L.~Espeholt, K.~Simonyan, A.~v.~d. Oord, A.~Graves, and
  K.~Kavukcuoglu, ``Neural machine translation in linear time,'' \emph{arXiv
  preprint arXiv:1610.10099}, 2016.

\bibitem{oord2016wavenet}
A.~v.~d. Oord, S.~Dieleman, H.~Zen, K.~Simonyan, O.~Vinyals, A.~Graves,
  N.~Kalchbrenner, A.~Senior, and K.~Kavukcuoglu, ``Wavenet: A generative model
  for raw audio,'' \emph{arXiv preprint arXiv:1609.03499}, 2016.

\bibitem{wang2019towards}
J.~Wang, Q.~Liu, Z.~Liu, and S.~Wu, ``Towards accurate and interpretable
  sequential prediction: A cnn \& attention-based feature extractor,'' in
  \emph{Proceedings of the 28th ACM International Conference on Information and
  Knowledge Management}, 2019, pp. 1703--1712.

\bibitem{yuan2020future}
F.~Yuan, X.~He, H.~Jiang, G.~Guo, J.~Xiong, Z.~Xu, and Y.~Xiong, ``Future data
  helps training: Modeling future contexts for session-based recommendation,''
  in \emph{Proceedings of The Web Conference 2020}, 2020, pp. 303--313.

\bibitem{sun2020generic}
Y.~Sun, F.~Yuan, M.~Yang, G.~Wei, Z.~Zhao, and D.~Liu, ``A generic network
  compression framework for sequential recommender systems,'' \emph{arXiv
  preprint arXiv:2004.13139}, 2020.

\bibitem{wang2020stackrec}
J.~Wang, F.~Yuan, J.~Chen, Q.~Wu, C.~Li, M.~Yang, Y.~Sun, and G.~Zhang,
  ``Stackrec: Efficient training of very deep sequential recommender models by
  layer stacking,'' \emph{arXiv preprint arXiv:2012.07598}, 2020.

\bibitem{chen2021user}
L.~Chen, F.~Yuan, J.~Yang, X.~Ao, C.~Li, and M.~Yang, ``A user-adaptive layer
  selection framework for very deep sequential recommender models,'' in
  \emph{AAAI}, 2021.

\bibitem{tan2016improved}
Y.~K. Tan, X.~Xu, and Y.~Liu, ``Improved recurrent neural networks for
  session-based recommendations,'' in \emph{DLRS}, 2016, pp. 17--22.

\bibitem{tang2018personalized}
J.~Tang and K.~Wang, ``Personalized top-n sequential recommendation via
  convolutional sequence embedding,'' in \emph{WSDM}.\hskip 1em plus 0.5em
  minus 0.4em\relax ACM, 2018, pp. 565--573.

\bibitem{sun2019bert4rec}
F.~Sun, J.~Liu, J.~Wu, C.~Pei, X.~Lin, W.~Ou, and P.~Jiang, ``Bert4rec:
  Sequential recommendation with bidirectional encoder representations from
  transformer,'' in \emph{Proceedings of the 28th ACM international conference
  on information and knowledge management}, 2019, pp. 1441--1450.

\bibitem{hidasi2018recurrent}
B.~Hidasi and A.~Karatzoglou, ``Recurrent neural networks with top-k gains for
  session-based recommendations,'' in \emph{CIKM}.\hskip 1em plus 0.5em minus
  0.4em\relax ACM, 2018, pp. 843--852.

\bibitem{gabriel2019contextual}
P.~M. Gabriel De~Souza, D.~Jannach, and A.~M. Da~Cunha, ``Contextual hybrid
  session-based news recommendation with recurrent neural networks,''
  \emph{IEEE Access}, vol.~7, pp. 169\,185--169\,203, 2019.

\bibitem{zeng2021knowledge}
Z.~Zeng, C.~Xiao, Y.~Yao, R.~Xie, Z.~Liu, F.~Lin, L.~Lin, and M.~Sun,
  ``Knowledge transfer via pre-training for recommendation: A review and
  prospect,'' \emph{Frontiers in big Data}, vol.~4, 2021.

\bibitem{khan2017cross}
M.~M. Khan, R.~Ibrahim, and I.~Ghani, ``Cross domain recommender systems: a
  systematic literature review,'' \emph{ACM Computing Surveys (CSUR)}, vol.~50,
  no.~3, pp. 1--34, 2017.

\bibitem{kanagawa2019cross}
H.~Kanagawa, H.~Kobayashi, N.~Shimizu, Y.~Tagami, and T.~Suzuki, ``Cross-domain
  recommendation via deep domain adaptation,'' in \emph{European Conference on
  Information Retrieval}.\hskip 1em plus 0.5em minus 0.4em\relax Springer,
  2019, pp. 20--29.

\bibitem{bousmalis2016domain}
K.~Bousmalis, G.~Trigeorgis, N.~Silberman, D.~Krishnan, and D.~Erhan, ``Domain
  separation networks,'' in \emph{Advances in neural information processing
  systems}, 2016, pp. 343--351.

\bibitem{yuan2019darec}
F.~Yuan, L.~Yao, and B.~Benatallah, ``Darec: Deep domain adaptation for
  cross-domain recommendation via transferring rating patterns,'' \emph{arXiv
  preprint arXiv:1905.10760}, 2019.

\bibitem{hu2018conet}
G.~Hu, Y.~Zhang, and Q.~Yang, ``Conet: Collaborative cross networks for
  cross-domain recommendation,'' in \emph{Proceedings of the 27th ACM
  International Conference on Information and Knowledge Management}, 2018, pp.
  667--676.

\bibitem{ni2018perceive}
Y.~Ni, D.~Ou, S.~Liu, X.~Li, W.~Ou, A.~Zeng, and L.~Si, ``Perceive your users
  in depth: Learning universal user representations from multiple e-commerce
  tasks,'' in \emph{KDD}, 2018, pp. 596--605.

\bibitem{yang2019xlnet}
Z.~Yang, Z.~Dai, Y.~Yang, J.~Carbonell, R.~R. Salakhutdinov, and Q.~V. Le,
  ``Xlnet: Generalized autoregressive pretraining for language understanding,''
  in \emph{Advances in neural information processing systems}, 2019, pp.
  5754--5764.

\bibitem{bertinetto2016learning}
L.~Bertinetto, J.~F. Henriques, J.~Valmadre, P.~Torr, and A.~Vedaldi,
  ``Learning feed-forward one-shot learners,'' in \emph{Advances in neural
  information processing systems}, 2016, pp. 523--531.

\bibitem{rendle2009bpr}
S.~Rendle, C.~Freudenthaler, Z.~Gantner, and L.~Schmidt-Thieme, ``Bpr: Bayesian
  personalized ranking from implicit feedback,'' in \emph{Proceedings of the
  twenty-fifth conference on uncertainty in artificial intelligence}.\hskip 1em
  plus 0.5em minus 0.4em\relax AUAI Press, 2009, pp. 452--461.

\bibitem{maddison2016concrete}
C.~J. Maddison, A.~Mnih, and Y.~W. Teh, ``The concrete distribution: A
  continuous relaxation of discrete random variables,'' \emph{arXiv preprint
  arXiv:1611.00712}, 2016.

\bibitem{jang2016categorical}
E.~Jang, S.~Gu, and B.~Poole, ``Categorical reparameterization with
  gumbel-softmax,'' \emph{arXiv preprint arXiv:1611.01144}, 2016.

\bibitem{wu2018blockdrop}
Z.~Wu, T.~Nagarajan, A.~Kumar, S.~Rennie, L.~S. Davis, K.~Grauman, and
  R.~Feris, ``Blockdrop: Dynamic inference paths in residual networks,'' in
  \emph{CVPR}, 2018, pp. 8817--8826.

\bibitem{guo2019spottune}
Y.~Guo, H.~Shi, A.~Kumar, K.~Grauman, T.~Rosing, and R.~Feris, ``Spottune:
  transfer learning through adaptive fine-tuning,'' in \emph{CVPR}, 2019, pp.
  4805--4814.

\bibitem{DBLP:conf/slsp/Bengio13}
Y.~Bengio, ``Deep learning of representations: Looking forward,'' in
  \emph{{SLSP}}, ser. Lecture Notes in Computer Science.\hskip 1em plus 0.5em
  minus 0.4em\relax Springer, 2013, pp. 1--37.

\bibitem{DBLP:conf/cvpr/RennieMMRG17}
S.~J. Rennie, E.~Marcheret, Y.~Mroueh, J.~Ross, and V.~Goel, ``Self-critical
  sequence training for image captioning,'' in \emph{{CVPR}}, 2017, pp.
  1179--1195.

\bibitem{williams1992simple}
R.~J. Williams, ``Simple statistical gradient-following algorithms for
  connectionist reinforcement learning,'' \emph{Machine learning}, pp.
  229--256, 1992.

\bibitem{long2015learning}
M.~Long, Y.~Cao, J.~Wang, and M.~I. Jordan, ``Learning transferable features
  with deep adaptation networks,'' \emph{arXiv preprint arXiv:1502.02791},
  2015.

\bibitem{caruana1997multitask}
R.~Caruana, ``Multitask learning,'' \emph{Machine learning}, vol.~28, no.~1,
  pp. 41--75, 1997.

\bibitem{he2018adversarial}
X.~He, Z.~He, X.~Du, and T.~Chua, ``Adversarial personalized ranking for
  recommendation,'' \emph{SIGIR}, pp. 355--364, 2018.

\end{thebibliography}
%



%
\begin{IEEEbiography}[{\includegraphics[width=1in,height=1.25in,clip,keepaspectratio]{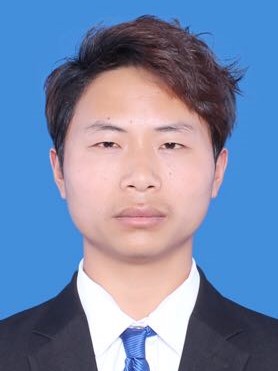}}]{Lei Chen}
is currently a master student at Shenzhen Institute of Advanced Technology, Chinese Academy of Sciences. He received the B.S. degree in computer science and technology from China University of Mining and Technology  in  2019.  His research interests  include  natural language processing, recommendation systems and information retrieval.
\end{IEEEbiography}

\begin{IEEEbiography}[{\includegraphics[width=1in,height=1.25in,clip,keepaspectratio]{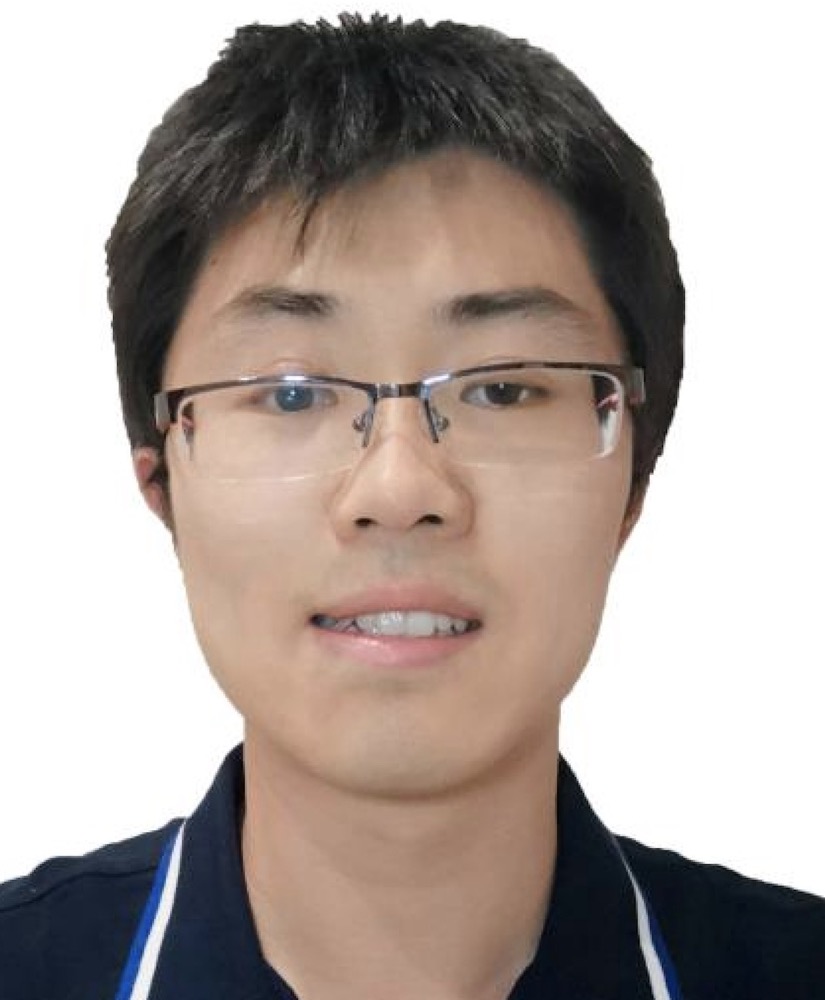}}]{Fajie Yuan}
is currently an assistant professor at Westlake University. Prior to that, he was a senior AI researcher
at Tencent working on recommender systems. He received his Ph.D. degree from the University of Glasgow.
 His main research interests include deep learning and transfer learning and their applications in recommender systems.
\end{IEEEbiography}

\begin{IEEEbiography}[{\includegraphics[width=1in,height=1.25in,clip,keepaspectratio]{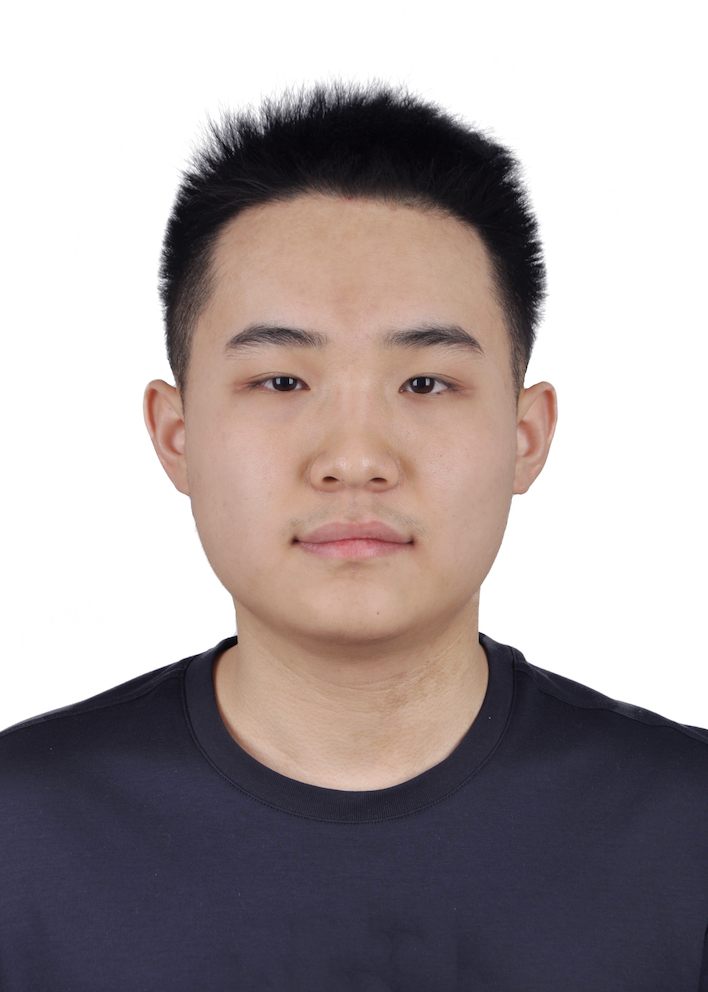}}]{Jiaxi Yang}
is currently an undergraduate of Huazhong University of Science and Technology. He will study for a Ph.D. degree at Shenzhen Institute of Advanced Technology, Chinese Academy of Sciences. His research interests include recommendation systems and information retrieval.
\end{IEEEbiography}

\begin{IEEEbiography}[{\includegraphics[width=1in,height=1.25in,clip,keepaspectratio]{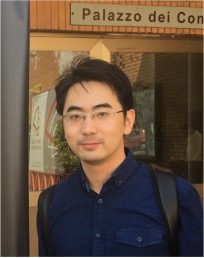}}]{Xiangnan He}
is a professor at the University of Science and Technology of China (USTC). He received his Ph.D. in Computer Science from the National University of Singapore (NUS). His research interests span information retrieval, data mining, and multi-media analytics. He has over 80 publications that appeared in several top conferences such as SIGIR, WWW, and MM, and journals including TKDE, TOIS, and TMM. His work has received the Best Paper Award Honorable Mention in WWW 2018 and ACM SIGIR 2016. He is in the editorial board of journals including Frontiers in Big Data, AI Open etc. Moreover, he has served as the PC chair of CCIS 2019 and SPC/PC member for several top conferences including SIGIR, WWW, KDD, MM, WSDM, ICML etc., and the regular reviewer for journals including TKDE, TOIS, TMM, etc.
\end{IEEEbiography}

\begin{IEEEbiography}[{\includegraphics[width=1in,height=1.25in,clip,keepaspectratio]{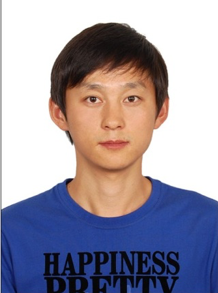}}]{Chengming Li}
received the B.S. and M.S. degree in computer application technology from Dalian University of Technology in 2009 and 2011, respectively. He received Ph.D. degree in Graduate School of Information Science and Electrical Engineering from Kyushu University in 2015. He currently is an associate professor at Shenzhen Institute of Advanced Technology, Chinese Academy of Sciences. His research interests include Data Mining, Network Security, and Big Data.
\end{IEEEbiography}

\begin{IEEEbiography}[{\includegraphics[width=1in,height=1.25in,clip,keepaspectratio]{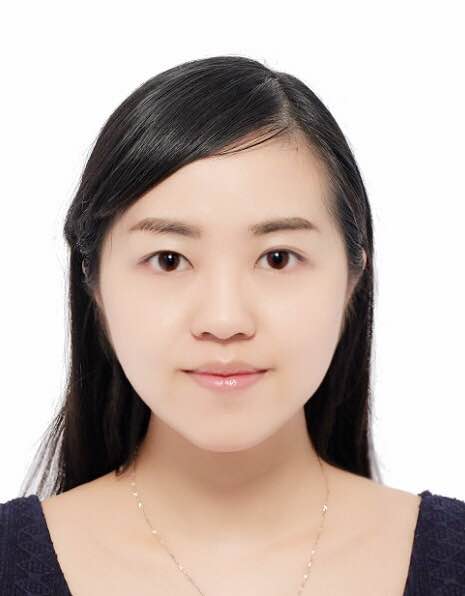}}]{Min Yang}
is currently an associate professor with the Shenzhen Institute of Advanced Technology, Chinese Academy of Science. She received her Ph.D. degree from the University of Hong Kong in February 2017. Prior to that, she received her B.S. degree from Sichuan University in 2012.  Her current research interests include recommendation systems, deep learning and natural language processing.
\end{IEEEbiography}









\end{document}